\documentclass[twocolumn,twocolappendix]{aastex6}

\def\farcs{\hbox{$.\!\!^{\prime\prime}$}}
\def\arcs{^{\prime\prime}}
\def\simgeq{{\raise.0ex\hbox{$\mathchar"013E$}\mkern-14mu\lower1.2ex\hbox{$\mathchar"0218$}}} 
\def\Sersic{S\'{e}rsic}
\def\H160{$H_{160}$}
\def\Re{$R_{\rm{e}}$}

\def\SN{$\rm{S/N}$}
\def\ML{$M/L^{\ast}$}
\def\MLh{$M/L^{\ast}_{\rm H}$}
\def\lMLh{$\log{(M/L^{\ast}_{\rm H})}$}
\def\mum{${\rm \mu}$m}
\def\a3{{${\rm A^3COSMOS}$}}

\def\rr{$R_{{\rm e,submm}}/R_{{\rm e,*}}$}
\def\rrd{$R_{{\rm e,dust}}/R_{ {\rm e,*}}$}

\def\at3{${\rm ATLAS^{3D}}$}
\def\mjb{$\rm{mJy \,beam^{-1}}$}
\usepackage{enumerate}
\usepackage{caption}
\usepackage{subcaption}
\usepackage{color}
\usepackage[bottom]{footmisc}

\collaborationName{Friends of AASTeX}

\begin{document}

\title{Revealing the stellar mass and dust distributions of submillimeter galaxies at redshift 2}

\author{P. Lang\altaffilmark{1}, E. Schinnerer\altaffilmark{1}, Ian Smail\altaffilmark{2}, U. Dudzevi\v{c}i\={u}t\.e\altaffilmark{2}, A.M. Swinbank\altaffilmark{2}, Daizhong Liu\altaffilmark{1}, S. K. Leslie\altaffilmark{1}, O. Almaini\altaffilmark{3}, Fang Xia An\altaffilmark{2,4}, F. Bertoldi\altaffilmark{5}, A. W. Blain\altaffilmark{6}, S. C. Chapman\altaffilmark{7}, Chian-Chou Chen\altaffilmark{8}, C. Conselice\altaffilmark{9}, E. A. Cooke\altaffilmark{2}, K. E. K. Coppin\altaffilmark{10}, J. S. Dunlop\altaffilmark{11}, D. Farrah\altaffilmark{12,13}, Y. Fudamoto\altaffilmark{14}, J. E. Geach\altaffilmark{10}, B. Gullberg\altaffilmark{2}, K. C. Harrington\altaffilmark{5}, J. A. Hodge\altaffilmark{15}, R. J. Ivison\altaffilmark{8,11}, E. F. Jim\'enez-Andrade\altaffilmark{5,16}, B. Magnelli\altaffilmark{5}, M. J. Micha\l{}owski\altaffilmark{17}, P. Oesch\altaffilmark{14}, D. Scott\altaffilmark{18}, J. M. Simpson\altaffilmark{19}, V. Smol\v{c}i\'{c}\altaffilmark{20}, S. M. Stach\altaffilmark{2}, A. P. Thomson\altaffilmark{21}, S. Toft\altaffilmark{22}, E. Vardoulaki\altaffilmark{5}, J. L. Wardlow\altaffilmark{23}, A. Weiss\altaffilmark{24}, P. van der Werf\altaffilmark{15}}

\altaffiltext{1}{Max-Planck-Institut f\"{u}r Astronomie, K\"{o}nigstuhl 17, D-69117 Heidelberg, Germany}
\altaffiltext{2}{Center for Extragalactic Astronomy, Department of Physics, Durham University, South Road, Durham DH1 3LE, UK}
\altaffiltext{3}{School of Physics and Astronomy, University of Nottingham, University Park, Nottingham NG7 2RD, UK}
\altaffiltext{4}{Department of Physics and Astronomy, University of the Western Cape, Robert Sobukwe Road, 7535}
\altaffiltext{5}{Argelander-Institut f\"{u}r Astronomie, Universit\"{a}t of Bonn, Auf dem H\"{u}gel 71, 53121 Bonn, Germany}
\altaffiltext{6}{Department of Physics and Astronomy, University of Leicester, University Road, Leicester LE1 7RH, UK}
\altaffiltext{7}{Department of Physics and Atmospheric Science, Dalhousie University Halifax, NS B3H 3J5, Canada}
\altaffiltext{8}{European Southern Observatory, Karl Schwarzschild Strasse 2, Garching, Germany}
\altaffiltext{9}{University of Nottingham, School of Physics and Astronomy, Nottingham, NG7 2RD, UK}
\altaffiltext{10}{Centre for Astrophysics Research, School of Physics, Astronomy and Mathematics, University of Hertfordshire, Hatfield AL10 9AB, UK}
\altaffiltext{11}{Institute for Astronomy, University of Edinburgh, Royal Observatory, Blackford Hill, Edinburgh EH9 3HJ, UK}
\altaffiltext{12}{Department of Physics and Astronomy, University of Hawaii, 2505 Correa Road, Honolulu, HI 96822, USA}
\altaffiltext{13}{Institute for Astronomy, 2680 Woodlawn Drive, University of Hawaii, Honolulu, HI 96822, USA}
\altaffiltext{14}{Department of Astronomy, University of Geneva, Ch. des Maillettes 51, 1290 Versoix, Switzerland}
\altaffiltext{15}{Leiden Observatory, Leiden University, P.O. Box 9513, 2300 RA Leiden, the Netherlands}
\altaffiltext{16}{International Max Planck Research School of Astronomy and Astrophysics at the Universities of Bonn and Cologne, Bonn, Germany}
\altaffiltext{17}{Astronomical Observatory Institute, Faculty of Physics, Adam Mickiewicz University, ul. S\l{}oneczna 36, 60-286 Pozna\'{n}, Poland}
\altaffiltext{18}{Department of Physics and Astronomy, University of British Columbia, 6224 Agricultural Road, Vancouver, BC V6T 1Z1, Canada}
\altaffiltext{19}{Academia Sinica Institute of Astronomy and Astrophysics, No. 1, Section 4, Roosevelt Road, Taipei 10617, Taiwan}
\altaffiltext{20}{Faculty of Science, University of Zagreb, Bijeni\u{c}ka c. 32, 10002 Zagreb, Croatia}
\altaffiltext{21}{The University of Manchester, Oxford Road, Manchester, M13 9PL, UK}
\altaffiltext{22}{Cosmic Dawn Center (DAWN), Niels Bohr Institute, University of Copenhagen, Lyngbyvej 2, DK-2100 Denmark}
\altaffiltext{23}{Physics Department, Lancaster University, Lancaster, LA1 4YB, UK}
\altaffiltext{24}{Max-Planck-Institut f\"{u}r Radioastronomie, Auf dem H\"{u}gel 69, D-53121 Bonn, Germany}

\begin{abstract}
We combine high-resolution ALMA and {\it HST}/CANDELS observations of 20 submillimeter galaxies (SMGs) predominantly from the AS2UDS survey at $z \simeq 2$ with bright rest-frame optical counterparts ($K_{\rm s} \lesssim 22.9$) to investigate the resolved structural properties of their dust and stellar components. We derive two-dimensional stellar-mass distributions that are inferred from spatial mass-to-light ratio (\ML) corrections based on rest-frame optical colors. Due to the high central column densities of dust in our SMGs, our mass distributions likely represent a lower limit to the true central mass density. The centroid positions between the inferred stellar-mass and the dust distributions agree within $1.1$\,kpc, indicating an overall good spatial agreement between the two components. The majority of our sources exhibit compact dust configurations relative to the stellar component (with a median ratio of effective radii \rrd\ {$=0.6$}). This ratio does not change with specific star-formation rate (sSFR) over the factor of 30 spanned by our targets, sampling the locus of `normal' main sequence galaxies up to the starburst regime, $\log{({\rm{\rm sSFR/sSFR_{MS}}})}\ge0.5$. Our results imply that massive SMGs are experiencing centrally enhanced star formation unlike typical spiral galaxies in the local Universe. The sizes and stellar densities of our SMGs are in agreement with those of the passive population at $z=1.5$, consistent with these systems being the descendants of $z\simeq2$ SMGs. 
\end{abstract}

\section{Introduction}

Numerical simulations suggest that the majority of actively star-forming galaxies during the peak epoch of galaxy formation grew through smooth accretion of cold gas triggering internal star formation processes \citep{Keres2005,Bournaud2009,Dekel2009}, without the need for major interactions. This basic picture has been inferred from a relatively tight relation between the star-formation rate and the assembled stellar mass, the `main sequence' of star-forming galaxies, \citep[][]{Noeske2007,Daddi2007,Elbaz2007,Sargent2014,Salmon2015,Speagle2014} which is claimed to exist out to redshifts around 6. Moreover, systematic studies of the ISM content, ionized gas kinematics and morphologies of star-forming galaxies at $z =1$-3 have revealed that a high fraction of these systems are large, gas-rich \citep[e.g.,\,][]{Tacconi2010,Tacconi2013,Daddi2010,Swinbank2012,Bothwell2013,Genzel2015,Schinnerer2016} rotating disks \citep[e.g.,\,][]{Forster2009,Law2009,Wisnioski2015,Swinbank2015,Swinbank2017} with exponential light and mass profiles \citep{Wuyts2011,Bell2012,Lang2014}. Bulge-to-disk decompositions have shown that high-redshift star-forming galaxies at the highest stellar masses ($\log{(M_*/{\rm M_{\sun}})} > 11$) exhibit significant stellar bulges, suggesting an evolutionary connection to today's massive quiescent ellipticals \citep[e.g.,\,][]{Lang2014,Bruce2014,Tacchella2015a,Tacchella2015b}.

At the highest star-formation rates, there is currently no observational or theoretical consensus on the triggering mechanism for star formation. The most luminous class of such systems are submillimeter galaxies \citep[SMGs, ][and references therein]{Blain2002,Casey2014}, which are mm/submm bright sources first detected in ground-based submm surveys. They are a class of strongly dust-obscured galaxies and are associated with large infrared luminosities ($L_{{\rm IR}} > 10^{12} \, {\rm L_{\sun}}$; implying extreme star-formation rates up to several $1000 \, {\rm M_{\sun} \, yr^{-1}}$), and are predominantly found at $z \sim 1 - 3$, with a substantial tail to $z \sim 6$ in their redshift distribution \citep[e.g.,\,][]{Smail1997,Chapman2005,Wardlow2011,Smolcic2012,Weiss2013,Simpson2014}. SMGs appear to be gas-rich systems with short depletion time scales of about 100 Myr \citep[e.g.,\,][]{Frayer1998,Frayer1999,Greve2005,Tacconi2008,Bothwell2013,Miettinen2017}. Since SMGs contribute 10-30\,\% of the total star-formation rate at $z=1$-4 \citep[e.g.,\,][]{Barger2012,Swinbank2014}, it is crucial to understand their connection to the rest of the star-forming galaxy population. Furthermore, the observed properties of the SMG population such as clustering, stellar masses, star-formation rates, gas masses and implied burst times are indicative of SMGs being progenitors of local early-type galaxies (ETGs) \citep[e.g.,\,][]{Hickox2012,Simpson2014,Bothwell2013}.

In the most basic picture, the infrared luminosities of SMGs suggest they are the high-redshift analoges of ultra-luminous infrared galaxies (ULIRGs; $L_{{\rm IR}} > 10^{12} \, {\rm L_{\sun}}$) observed in the local Universe \citep[][ for a review]{Sanders1996,Londsdale2006}. Local ULIRGs are typically triggered through major galaxy mergers \citep[e.g.,\,][]{Sanders1988}, and some galaxy evolution models propose a similar origin for SMGs \citep[e.g.,\,][]{Narayanan2010}. Alternative models suggest that SMGs are massive, isolated disk galaxies with strong secular bursts of star formation \citep[e.g.,\,][]{Dave2010}. Finally, SMGs might be comprised of a heterogeneous population of both merger-induced systems and secularly evolving disk galaxies \citep{Hayward2011,Hayward2013}.

One key element to distinguish between the proposed evolutionary scenarios, is to investigate the spatial distribution of their ongoing star formation and existing stellar components. Tracing the rest-frame far-infrared continuum emission (i.e.,\,corresponding to observed submillimeter wavelengths) with instruments such as the Atacama Large Millimeter/submillimeter Array (ALMA), has enabled systematic studies of the resolved distribution of star formation in significant samples of high-redshift SMGs. 

These studies conclude that obscured star formation is confined to compact, central components with half-light radii of 1-2\,kpc \citep[e.g.,\,][]{Swinbank2010a,Ikarashi2015,Simpson2015b,Hodge2016,Oteo2017,Gullberg2018}. Determining the radial structure of the resolved far-infrared emission, \cite{Hodge2016,Hodge2018} found that the far-infrared emission in SMGs at $z\simeq 2.5$ is consistent with 
exponential profiles without strong evidence for clumpy disk structures. However, the occurrence of these compact star-forming disks is not sufficient to constrain the formation mechanisms of SMGs. For example, they might be triggered by the dissipative collapse of gas during major galaxy mergers \citep[e.g.,\,][]{Bournaud2011}, or radial flows of gas caused by disk instabilities might trigger strong centrally enhanced star formation within massive star-forming disks without the need for on-going merging \citep[e.g.,\,][]{Elmegreen2008,Dekel2014,Bournaud2014}.
In parallel to studying the far-infrared component, stellar morphologies of SMGs have also been investigated by probing their resolved rest-frame optical emission. Such studies have revealed that SMGs have disturbed and irregular morphologies, with exponential radial profiles \citep[][]{Swinbank2010b,Targett2013,Chen2015,Hodge2016,Rujopakarn2016}. Although these properties are similar to those of the `normal' star-forming galaxy population, signatures of tidal tails and asymmetric structures indicative of mergers are also more frequently found in SMGs or far-infrared selected galaxies at $z \ge 1$, in contrast to their counterparts at lower star-formation rates \citep{Conselice2011,Wiklind2014,Kartaltepe2012}. Moreover, recent studies combined {\it HST} imaging with ALMA observations (albeit based on small samples) to test whether the compact far-infrared emission is spatially decoupled from the more extended optical emission, and hence to determine if this arrangement is in agreement with a merger origin of SMGs \citep[e.g.,\,][]{Ivison2000,Chen2015,Hodge2016}.

One major caveat of these studies is that the optical emission can be strongly affected by structured dust extinction \citep[see e.g.,\,][]{Simpson2017} , and thus might fail to be a reliable tracer of the underlying stellar-mass distribution. In the most extreme cases, inferred column densities in the centers of high-redshift SMGs imply optical extinctions reaching $A_{\rm V} \sim 2000$ \citep{Gomez2018,Simpson2017}. In such cases, the observed optical emission can clearly not be used to infer the underlying stellar-mass distribution. 
However, based on samples of optically-selected star-forming galaxies at $1 < z < 3$, it is possible to recover the underlying stellar mass distributions of galaxies and search for radial gradients in color and hence \ML\ \citep{Wuyts2012,Lang2014}.

Here, we extend these studies to investigate the formation mechanisms of SMGs by combining information on the resolved distribution of both the dust-obscured star formation and the assembled stellar mass on kpc scales. We analyze dust continuum imaging at $870\,\mu$m of 19 SMGs around $z$=2 from three cosmological deep fields. Reconstructed stellar mass distributions at kpc-scale resolution are available from the deep multi-wavelength imaging from the CANDELS survey \citep{Koekemoer2011,Grogin2011}. Through analyzing this unique data set of submm/mm-selected SMGs probing a large range of star-formation rates ($ 100\, {\rm M_{\sun} \, yr^{-1}} \lesssim {\rm SFR} \lesssim 700\, {\rm M_{\sun} \, yr^{-1}}$), we compare the dust-obscured star formation distribution to the inferred stellar morphologies. This analysis will help to elucidate the evolutionary connections of distant SMGs to less actively star-forming galaxies at similar redshifts, as well as to the local population of massive spheroids.

Our paper is arranged as follows: We discuss the various sources of observational data used, as well as our sample selection in Section 2. This is followed by a description of our methodologies to derive stellar-mass maps and morphological quantities in Section 3. Our results are presented in Section 4, followed by a discussion of our findings in Section 5. Finally, our conclusions are presented in Section 6. Throughout this paper, we assume a \cite{Chabrier2003} initial mass function (IMF), and adopt the cosmological parameters $(\Omega _M, \Omega _{\Lambda}, h) = (0.3, 0.7, 0.7)$. 

\section{Observations and sample}

To study the structure of stars and obscured star formation in SMGs, we construct a sample of 20 SMGs that have been targeted with deep, high angular-resolution observations from both ALMA and {\it HST} across three cosmological deep fields.

As a basis for our sample, we consider SMGs originally detected in ground-based single-dish submm surveys and followed-up with high-resolution ALMA observations in the submm continuum. 
We apply the following selection criteria (see Section\ \ref{final_sample.sec} for a more detailed discussion): (a)
redshift range $1.7 < z < 2.6$ (where our method to derive stellar-mass maps is most robust), (b) coverage by the {\it HST}/CANDELS survey, and that are detected with sufficient S/N in both $J_{125}$ and $H_{160}$ band filters, and (c) no evidence for an AGN.

As high-resolution ALMA and {\it HST} observations with sufficient S/N are only available for a limited number of SMGs, this requirement imposes the strongest limit on our sample size and prevents the use of further selection criteria (such as, e.g., flux-limits). 
We discuss the properties of our sample and potential sample biases due to our various selection criteria in Section\ \ref{final_sample.sec}.

In the remainder of this section, we describe the ALMA observational programs from which data are used, the {\it HST} imaging, the ancillary multi-wavelength photometry, and the selection of our final sample. 

\subsection{ALMA observations}

\subsubsection{UDS}

14 SMGs of our final sample are taken from the ALMA-SCUBA-2 Ultra Deep Survey (UDS), henceforth referred to as `AS2UDS'. AS2UDS is a follow-up study of a complete sample of 716 SCUBA2-sources in the UDS fields detected at $850\,\mu$m with ${\rm S/N}>4$ \citep[see][]{Geach2017}. 30 of the brightest sources were observed in ALMA Cycle 1 \citep[Project ID:2012.1.00090.S; ][]{Simpson2015b,Simpson2017} and the remaining 689 sources were observed in ALMA Cycle 3, 4, and 5 (Project IDs: 2015.1.01528.S, 2016.1.00434.S, and 2017.1.01492.S, respectively) at $870\,\mu$m. The final detection maps have median depths of $\sigma_{870}=0.25\,$\mjb\ (Cycle 1), $\sigma_{870}=0.34\,$\mjb\ (Cycle 3), $\sigma_{870}=0.23\,$\mjb (Cycle 4), and $\sigma_{870}=0.085\,$\mjb (Cycle 5). The resulting sample contains 708 individual ALMA sources with $S_{870}>0.6\,{\rm mJy}$ (corresponding to $4.3\sigma$). A full description can be found in \cite[][]{Stach2018,Stach2019,Simpson2015a,Simpson2015b}.

Here, we consider the subset of 696 AS2UDS SMGs with $S_{870}>1\,{\rm mJy}$, all for which photometric and/or spectroscopic redshifts are available. The ALMA images for AS2UDS sources used in this analysis have an angular resolution of FWHM $= 0\farcs19$-$ 0\farcs35$. Deboosted and primary beam-corrected flux densities are taken from \cite[][]{Stach2018}. Available $K_{\rm s}$-band magnitudes are taken from the UKIDSS UDS DR11 photometric catalog based on a median $3\sigma$ depth of 25.7 mag.

\subsubsection{ECDFS}

Another part of our sample is taken from the ALMA Band 7 follow-up of single-dish submm sources from the LESS survey \citep[`ALESS',\,][]{Hodge2013,Karim2013}, which provides a homogeneous and unbiased sample of SMGs over a wide redshift range in the ECDFS field (including GOODS-South). 126 LESS sources detected with ${\rm S/N}>3.7$ were observed in ALMA Cycle 0 \citep[Project ID: 2011.0.00294\,][]{Hodge2013} at $870\,\mu$m with a spatial resolution of FWHM=$1\farcs6$. The resulting sample of ALESS SMGs \citep[referred to as the `Main sample' in ][]{Hodge2013} comprises 99 SMGs detected in the ALMA maps above $3.5 \sigma$ (the ALMA maps have a median of $\sigma_{870}=0.4\,$\mjb).

A sub-sample were followed-up in ALMA Cycle 1 at Band 7 by \cite{Hodge2016}, providing high-resolution imaging (FWHM $\sim 0\farcs16$) at $870\,\mu$m for sixteen detected SMGs (ALMA project 2012.1.00307.S). For our study, we consider the whole set of 99 ALESS SMGs, for which photometric and/or spectroscopic redshifts are available. The ALMA imaging for ALESS sources is taken from Cycle 0 observations \cite{Hodge2013}. We supplemented these with the high-resolution Cycle 1 observation from \cite{Hodge2016} for one source (ALESS067.1). Deboosted and primary beam-corrected flux densities are taken from \cite{Hodge2013}. We take $K_{\rm s}$-band magnitudes for the ALESS sources from \cite{Simpson2014}, who combine $K_{\rm s}$ photometry from multiple surveys in the ECDFS field with $3\sigma$ limiting depths ranging from 22.4 to 24.4 mag. 

\subsubsection{COSMOS}
\label{a3.sec}

Additionally, we use the sample of SMGs detected with the ALMA Band 6 ($1.3\,$mm) follow-up of bright AzTEC sources in the COSMOS field. AzTEC SMGs were originally selected based on the $1.1\,$mm blank-field continuum survey with the ASTE/AzTEC instrument within COSMOS \citep{Aretxaga2011}. The 122 brightest AzTEC sources (${\rm S/N} > 4$) have been followed-up by ALMA observations at $1.3\,$mm with an angular resolution of FWHM=$1\farcs6$, and an r.m.s noise of $0.1 \, \rm{mJy \,beam^{-1}}$ (ALMA project 2013.1.00118.S, PI: M.Aravena). In total, 152 ALMA sources have been detected (with ${\rm S/N} \geq 5$) with ALMA. Out of those, 124 individual SMGs have been selected that have robust optical/near-infrared counterparts and thus photometric and/or spectroscopic redshifts. Furthermore, sources hosting AGN
have been removed by applying several criteria (based on detected X-ray as well as radio emission). Details of this catalog are presented in \cite{Miettinen2017}. For our study, we consider this set of 124 SMGs. The imaging for all our AzTEC SMGs is taken from the \a3 archive project that produces cleaned images for all publicly available ALMA continuum observations in COSMOS (see Liu et al. 2019 submitted for details). Deboosted and primary beam-corrected flux densities are taken from \cite{Miettinen2017}. $K_{\rm s}$-band magnitudes for the AzTEC sample are taken from the \cite{Brisbin2017} catalog, and correspond to a $3\sigma$ limiting depth of 24.0-24.7 mag \citep[for details see also ][]{Laigle2016}.

To compare the flux densities for all AzTEC SMGs observed at $1.3\,$mm with those from AS2UDS and ALESS (done in the following Section), we convert the flux density to observed-frame $870\,$\mum\ by applying the Rayleigh-Jeans approximation and assuming a dust emissivity index of $\beta = 2$. The choice of $\beta$ is motivated by the measurements from \cite{Magnelli2012}. We note that our conclusions are not significantly affected if we instead adopt $\beta = 1.5$.

\subsection{HST imaging}

For our study, we exploit {\it HST} observations from the deep multi-orbit CANDELS Treasury Survey \citep{Koekemoer2011,Grogin2011}. The CANDELS imaging covers multiple pass-bands at optical and near-infrared wavelengths from different cameras on board {\it HST}, from which we use the $J_{125}$ and $H_{160}$ filters observed with the WFC3 instrument.

All CANDELS imaging used in our analysis has been drizzled to a $0\farcs06$ pixel scale. 
Typical point-source limiting depths within CANDELS in \H160 are 27.0 mag. For details on the observations and data reduction, we refer the reader to \cite{Koekemoer2011} and \cite{Grogin2011}.

We have furthermore calibrated the astrometry of the \emph{HST} images for our high-resolution ALMA sample, and aligned the $H_{160}$-band images to the \emph{Spitzer}\,/\,IRAC 3.6-$\mu$m image of the full ECDFS and UDS fields. We used SExtractor to create a source catalog for each \emph{HST} image and the 3.6-$\mu$m image and match each source in the \emph{HST} catalog to the 3.6-$\mu$m catalog and measure an individual offset for each image. We apply a median offset of $\Delta$RA=$0\farcs{13}$ and $\Delta$Dec=$-$$0\farcs{27}$. We test the accuracy of our astrometry statistically by calculating the offsets between randomly chosen 3.6-$\mu$m sources and their $H_{160}$-band counterparts lying within a $0\farcs{8}$ radius (equivalent to the IRAC\,/\,3.6-$\mu$m point-spread function), and we find no systematic offset and a scatter of just $0\farcs{1}$ in both RA and Dec, consistent with the expected accuracy of the IRAC imaging \citep{Damen2011}.

\subsection{Ancillary data and integrated galaxy properties}

Integrated stellar masses and star-formation rates for the sources in our sample are derived using the {\sc Magphys} code \citep{daCunha2008}. {\sc Magphys} combines the emission from stellar populations with dust attenuation within galaxies by assuming energy balance. Spectral population synthesis models of \cite{Bruzual2003} are used in combination with a \cite{Chabrier2003} initial mass function. For this work, SED fitting was carried out with an updated version of {\sc Magphys} \citep{daCunha2015}, which includes updated recipes suited for galaxies at redshifts $>1$. Star formation histories (SFHs) are parametrized as delayed $\tau$ models with superimposed bursts of finite (30 to 300 Myr) length. Attenuation by dust in {\sc Magphys} is based on the model by \cite{Charlot2000}. \footnote{We note that recent measurements of the ${\rm ^{13}C/^{18}O}$ abundance ratio in a few SMGs at redshift $\simeq$ 2-3 indicate the presence of a top-heavy IMF \citep{Zhang2018}. Such IMF variations (alongside with various assumptions within SED-modeling concerning, e.g., the shape of the star formation history) might lead substantial systematic uncertainties in the derived stellar masses and star-formation rates of our sample (see Section\ \ref{final_sample_prop.sec}).}

The SED-derived parameters such as photometric redshifts, integrated star-formation rates, and stellar masses of our SMGs are taken from several studies in the literature which are in turn based on ground- and space-based photometry in ECDFS, COSMOS, and UDS:

For all AS2UDS SMGs, fits with {\sc Magphys} are taken from Dudzeviciute et al. (in prep). Briefly, the SEDs are modeled using photometry from U-band to MIPS 24\,$\mu$m together with deblended {\it Herschel} SPIRE data \citep[using the same technique as presented in\,][]{Swinbank2014}, ALMA Band 7 data and VLA $1.4\,$GHz radio data. Only a small subset (44 out of 695) of AS2UDS galaxies currently have spectroscopic redshifts, which we use for our analysis. 

For our ALESS SMGs, {\sc Magphys} parameters are available from \cite{daCunha2015} that rely on the determination of photometric redshifts. However, recent spectroscopic redshifts by \cite{Danielson2017} are now available for a subset of ALESS SMGs. As all ALESS SMGs in our final sample have a spectroscopic redshifts, we re-fit their SEDs with {\sc Magphys} including all ground- and space-based photometry as done in \cite{daCunha2015}, adopting the respective spectroscopic redshift.

Stellar properties for our AzTEC sources are taken from \cite{Miettinen2017} who employ SED-modeling with {\sc Magphys}. The photometry is taken from the COSMOS2015 catalog \citep{Laigle2016}, covering optical $B$-band to 24\,\mum\ fluxes. Additionally, de-blended far-infrared photometry (100-500\,\mum) from {\it Herschel}/PACS \& SPIRE \citep[see also\,][]{Brisbin2017} were included, as well as further ground-based far-infrared and radio measurements. We take the redshift information from \cite{Brisbin2017}, which compiles photometric redshifts and spectroscopic redshifts for 30 AzTEC SMGs.

We note that even for SMGs with spectroscopic redshifts, the detected ALMA/submm source might represent a lensed systems at higher redshifts that is not physically associated with its optical counterpart. The probability for a configuration of such lensed systems increases with submm flux of the source and reaches about $10\,\%$ at $S_{870} = 10\,$mJy \cite[i.e.,\,the maximum submm flux reached by our sample; ][]{Negrello2007}.

\subsection{Sample selection}
\label{final_sample.sec}

To construct our final SMG sample from the pool of available sources from the AS2UDS, ALESS, and AzTEC surveys (resulting in 919 sources), we apply the following selection criteria:

\begin{enumerate}

\item First, we consider all SMGs from these surveys that lie within $1.7 < z < 2.6$. At these redshifts, the reddest available {\it HST} imaging filters $J_{125}$ and $H_{160}$ (at a central wavelengths of 1.25 and 1.6\,\mum, respectively) sample the spectrum of a galaxy close to the Balmer break at $\sim 3800$\,\AA{}, while the $H_{160}$ falls redward of it. Therefore, this filter combination is well suited to derive stellar mass distributions for our sample sources. This reduces our initial sample to 362 galaxies. 

\item Next, we select sources that fall within the respective CANDELS areas of the UDS, GOODS-S, and COSMOS fields. Due to the limited overlap of CANDELS compared to the whole area covered by the respective deep fields, this criterion removes a large portion of the observed SMGs, leaving 26 galaxies.

\item We require a sufficient ${\rm S/N}$ level for the CANDELS imaging in both $J_{125}$ and $H_{160}$ filters to derive spatially resolved color and stellar mass distributions with the applied techniques (see Section\ \ref{color.ref} for a detailed explanation). In particular, we require enough ${\rm S/N}$ in both the $J_{125}$ and $H_{160}$ filters so that our color and hence \ML\ maps contain at least three spatial Voronoi bins to recover spatial varying \ML\ (see details in Section\ \ref{color.ref}). This criterion removes 6 galaxies with the faintest optical counterparts ($K_{\rm s} \gtrsim 22.6$ mag), such that 20 galaxies remain. 

\item Finally, we reject AGN based on various criteria. Since several AGN rejection criteria have already been applied to the parent samples from ALESS and AzTEC, we additionally reject AS2UDS galaxies that are either classified as AGN due to the \cite{Donley2012} criterion or have X-ray counterparts in the deep {\it Chandra} X-UDS catalog \citep{Kocevski2018}. One object from AS2UDS fulfills the AGN criteria (AS2UDS.292.0) through a close X-ray counterpart ($<1\arcs$) and is thus rejected from the sample.

\end{enumerate}

After applying these selection criteria, a total of 20 SMGs are considered for the final sample, from which 14, $3$, and $3$ are taken from AS2UDS, ALESS, and \a3/AzTEC, respectively. 

\subsection{Final sample properties}
\label{final_sample_prop.sec}
The final sample properties are provided in Table\,\ref{Sample.tbl}. Our sample has a median redshift and scatter of $z=2.15$ and 0.26, respectively. 16 galaxies out of our final sample have spectroscopic redshift information. For $14$ galaxies, our ALMA observations reach a resolution of better than $0\farcs2$, which is comparable to the {\it HST}/WFC3 imaging. Although being observed at slightly lower resolution ($0\farcs35$), we include AS2UDS.583.0 and AS2UDS.659.0 also in this high-resolution sample. These targets form a crucial sub-sample ideally suited to compare their stellar and far-infrared profiles at the same angular resolution. We keep our SMGs observed with ALMA at lower angular resolution (ranging from $0\farcs8$ to $2\farcs1$) to analyze their stellar mass distributions. Within the uncertainties, the median redshift, stellar mass and star-formation rate of the high-resolution sample do not change with respect to our full SMG sample.

\begin{table*}[t]
{\small
\centering
\caption[a]{Sample properties of our final SMG sample.}
\begin{tabular}{l*{8}{c}r}
\hline
\hline

ID$^a$ & $z^b$  &Band$^c$ & $S_{{\rm 870\,\mu m}}$$^d$ &Beam FWHM$^e$ & $K_{\rm s}$ & $H_{160}$$^f$ & $\log(M_*)^g$ & SFR$^h$ \\
  &   &  &  [mJy] & $[^{\prime\prime}]$ &[AB mag] &[AB mag] & [${\rm M_{\sun}}$] & [${\rm M_{\sun} \, yr^{-1}}$] \\
\hline
AS2UDS.113.1   &  $1.682$                  & 7  &  $2.9 \pm 0.3$   & 0.19 & $22.73\pm  0.02$ &  $23.15\pm 0.06 $ &  $10.30^{+0.11}_{-0.12}  $     &  $196^{+3}_{-3}$    \\
AS2UDS.116.0   &  $2.222$                  & 7  &  $6.0 \pm 0.6$   & 0.20 & $21.54\pm  0.01$ &  $22.24\pm 0.03 $ &  $11.62^{+0.11}_{-0.12}  $     &  $96^{+41}_{-34}$  \\
AS2UDS.125.0   &  $2.154$                  & 7  &  $4.6 \pm 0.5$   & 0.20 & $20.76\pm  0.01$ &  $21.21\pm 0.01 $ &  $11.70^{+0.11}_{-0.12}  $     &  $289^{+4}_{-0}$    \\
AS2UDS.153.0   &  $2.315$                  & 7  &  $3.2 \pm 0.5$   & 0.20 & $22.35\pm  0.02$ &  $22.37\pm 0.03 $ &  $10.72^{+0.06}_{-0.02}  $     &  $369^{+5}_{-70}$   \\
AS2UDS.259.0   &  $1.793$                  & 7  &  $4.7 \pm 0.3$   & 0.18 & $21.19\pm  0.01$ &  $21.79\pm 0.02 $ &  $11.29^{+0.11}_{-0.12}  $     &  $154^{+0}_{-2}$    \\
AS2UDS.266.0   &  $2.232$                  & 7  &  $4.2 \pm 0.7$   & 0.19 & $22.59\pm  0.02$ &  $23.29\pm 0.06 $ &  $11.01^{+0.04}_{-0.05}  $     &  $93^{+19}_{-18}$   \\
AS2UDS.271.0   &  $2.578$                  & 7  &  $3.9 \pm 0.7$   & 0.19 & $22.16\pm  0.01$ &  $22.52\pm 0.03 $ &  $10.93^{+0.04}_{-0.05}  $     &  $360^{+104}_{-51}$ \\
AS2UDS.272.0   &  $1.849$                  & 7  &  $5.1 \pm 0.5$   & 0.20 & $21.61\pm  0.01$ &  $22.13\pm 0.02 $ &  $11.11^{+0.13}_{-0.01}  $     &  $286^{+28}_{-205}$ \\
AS2UDS.297.0   &  $2.154$                  & 7  &  $4.4 \pm 0.6$   & 0.20 & $21.92\pm  0.01$ &  $22.82\pm 0.05 $ &  $10.69^{+0.01}_{-0.00}  $     &  $565^{+242}_{-195}$\\
AS2UDS.311.0   &  $1.995$                  & 7  &  $5.8 \pm 0.8$   & 0.19 & $21.76\pm  0.01$ &  $22.64\pm 0.04 $ &  $11.63^{+0.08}_{-0.07}  $     &  $142^{+28}_{-28}$  \\
AS2UDS.322.0   &  $2.542$                  & 7  &  $1.6 \pm 0.1$   & 0.80 & $21.98\pm  0.01$ &  $22.32\pm 0.03 $ &  $11.73^{+0.01}_{-0.03}  $     &  $171^{+2}_{-35}$  \\
AS2UDS.412.0   &  $2.450$                  & 7  &  $4.1 \pm 0.3$   & 0.18 & $22.51\pm  0.02$ &  $23.38\pm 0.08 $ &  $11.13^{+0.08}_{-0.03}  $     &  $207^{+53}_{-34}$  \\
AS2UDS.583.0   &  $2.47^{+0.16}_{-0.25}$   & 7  &  $3.1 \pm 0.4$   & 0.35 & $22.85\pm  0.03$ &  $23.09\pm 0.08 $ &  $10.51^{+0.12}_{-0.11}  $     &  $132^{+27}_{-27}$   \\
AS2UDS.659.0   &  $1.92^{+0.15}_{-0.17}$   & 7  &  $1.7 \pm 0.3$   & 0.35 & $21.60\pm  0.01$ &  $22.47\pm 0.03 $ &  $11.47^{+0.08}_{-0.09}  $     &  $114^{+23}_{-20}$  \\
ALESS018.1  &  $2.252$                     & 7  &  $4.4 \pm 0.5$   & 2.07 & $21.13\pm  0.01$ &  $22.01\pm 0.02 $ &  $11.47^{+0.05}_{-0.06}  $     &  $545^{+95}_{-65}$  \\
ALESS067.1  &  $2.123$                     & 7  &  $4.5 \pm 0.4$   & 0.18 & $21.09\pm  0.02$ &  $21.78\pm 0.02 $ &  $11.22^{+0.01}_{-0.01}  $     &  $154^{+2}_{-2}$  \\
ALESS079.2  &  $1.769$                     & 7  &  $2.0 \pm 0.4$   & 1.38 & $20.89\pm  0.01$ &  $21.53\pm 0.02 $ &  $11.43^{+0.01}_{-0.01}  $     &  $125^{+1}_{-1}$  \\
AzTECC33a   &  $2.30^{+0.46}_{-0.16}$      & 6  &  $7.0 \pm 0.7$   & 1.32 & $21.00\pm  0.10$ &  $21.47\pm 0.02 $ &  $10.99^{+0.01}_{-0.01}  $     &  $661^{+15}_{-1}$   \\
AzTECC38    &  $1.91^{+0.53}_{-0.46}$      & 6  &  $10.9\pm 0.6$   & 1.30 & $22.50\pm  0.10$ &  $23.42\pm 0.08 $ &  $11.52^{+0.01}_{-0.02}  $     &  $283^{+7}_{-13}$   \\
AzTECC95    &  $2.102$                     & 6  &  $5.3 \pm 0.4$   & 1.26 & $20.70\pm  0.10$ &  $21.39\pm 0.01 $ &  $11.28^{+0.01}_{-0.01}  $     &  $357^{+1}_{-1}$    \\

\hline
\hline
\end{tabular}
\caption*{{\bf Notes:} a) Source ID as adopted from \cite{Stach2018}, \cite{Hodge2013}, and \cite{Miettinen2017} for sources taken from AS2UDS, ALESS, and AzTEC samples, respectively. b) Source redshift. Values with quoted errors are photometric redshifts, and spectroscopic redshifts otherwise. c) Observed ALMA Band. d) Integrated total flux density at $870\,\mu$m. For sources observed in ALMA Band 6, fluxes were converted to $870\,\mu$m by applying the
Rayleigh-Jeans approximation and assuming a dust emissivity index of $\beta = 2$. e) Major axis beam FWHM size of the ALMA imaging. f) Integrated $H_{160}$ magnitude. g) and h) SED-derived galaxy-integrated stellar mass and star-formation rate derived from their optical+IR SEDs based on {\sc Magphys}.}
\label{Sample.tbl}}
\end{table*}

Our final sample of 20 sources represents only a small fraction of our SMG parent sample from AS2UDS, ALESS and AzTEC at $1.7 < z < 2.6$ (i.e.,\,our sample of SMGs that remain after applying selection criterion 1). Due to different flux limits of the parent surveys and the requirement of available redshifts, the selection of sub-samples for ALMA follow-up is complex. However, we treat these in the following as an indicative representation of the underlying SMG population with near-infrared counterparts at the flux limits applied, and test if our selection criteria might introduce selection biases. To do this, we explore the properties of our sample and compare those to the parent samples AS2UDS, ALESS and AzTEC at $1.7 < z < 2.6$ (i.e.,\,our sample of SMGs that remain after applying selection criterion 1). 

\begin {figure}[tb]
\centering
 \includegraphics[width=0.49\textwidth]{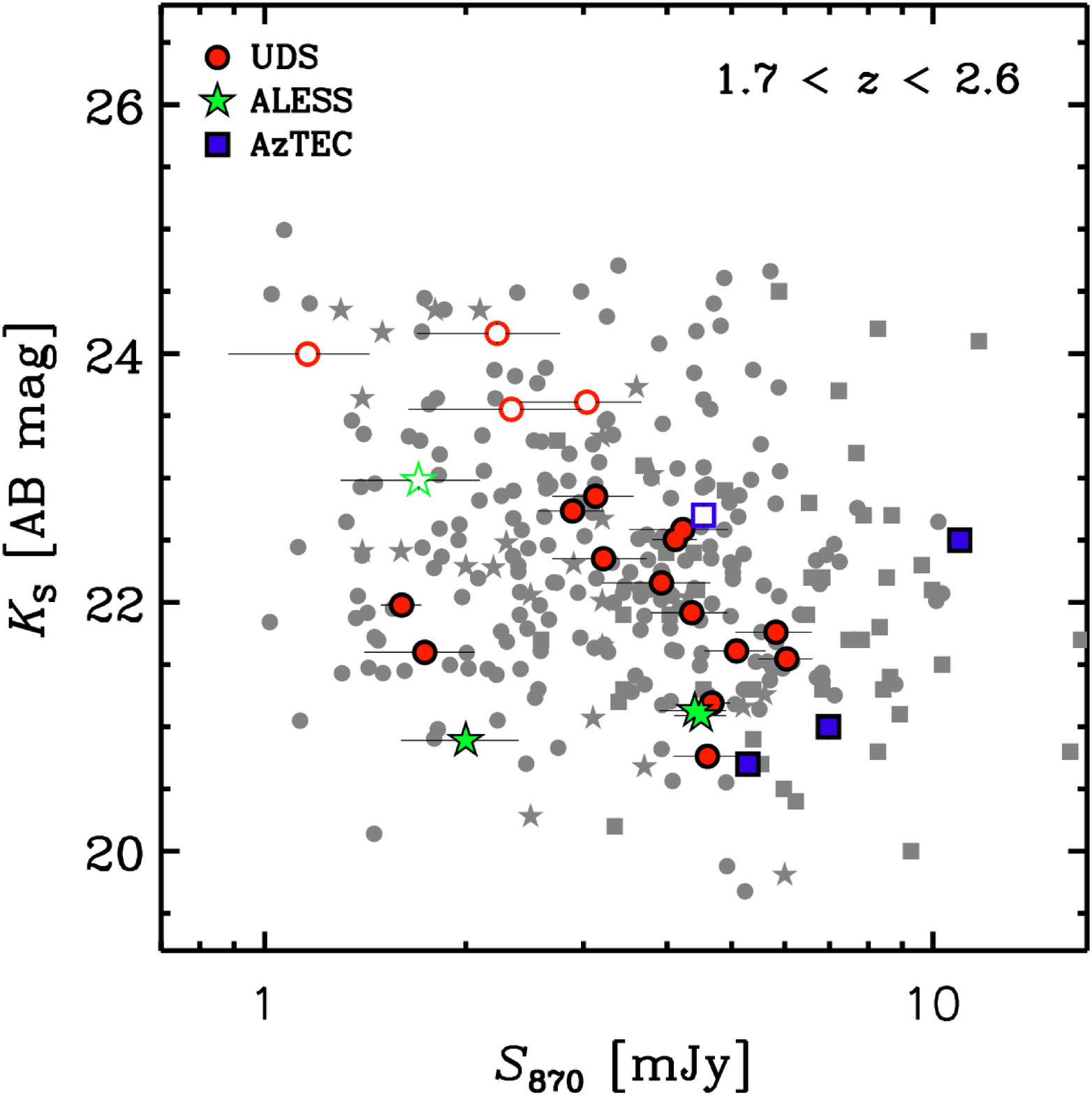}
\caption[s]{$K_{\rm s}$-band magnitude versus observed ALMA flux density at $870\,$\mum. Gray symbols show the parent samples of all AS2UDS, ALESS, and AzTEC SMGs in the redshift range $1.7 < z < 2.6$. Filled colored symbols represent the final SMG sample considered in this study. In addition, open colored symbols identify targets which are rejected due to their low surface brightness in the {\it HST}/WFC3 $J$ and $H$-band imaging. Our final selection yields optically/near-infrared bright SMGs ($K_{\rm s} \lesssim 22.9$) which overlap well with the underlying SMG parent samples with available redshifts.}

\label{Sample.fig}
 \vspace{3mm}
\end {figure}

In Figure\ \ref{Sample.fig}, we plot the apparent $K_{\rm s}$-band magnitude versus the observed flux density at $870\,$\mum\ ($S_{870}$) for the parent samples, and our {\it HST}/ALMA sample analyzed here after applying our redshift cut. We highlight all SMGs that fall within the CANDELS areas (i.e.,\,after applying criterion 2) as colored symbols. Those are split into faint near-infrared sources rejected by criterion 3 (i.e.,\,due to being too faint in either the $J_{125}$ and $H_{160}$ band for our analysis; open symbols) and our final SMG sample (filled symbols).

All SMGs that are covered by CANDELS show a fairly uniform and homogeneous sampling of the underlying SMG population at $z \simeq 2$, as the criterion of targets to fall within the CANDELS areas yield a random selection. 
Applying a Kolmogorov-Smirnov (K-S) test, we find that the distributions of both $K_{\rm s}$-band magnitude and $S_{870}$ are not significantly ($ \geq 3\sigma$) different when comparing the sources that are covered by CANDELS to the underlying SMG population. However, when comparing the distribution of our final sample and the sources too faint in either the $J_{125}$ and $H_{160}$ band, we find that all rejected SMGs have faint optical/near-infrared counterparts ($K_{\rm s} \gtrsim 22.6$ mag) and occupy the locus of low $870\,$\mum\ flux. As a consequence, our final sample exhibits a bias towards optically bright SMGs with high flux densities at $870\,\mu$m. The bias in $K_{\rm s}$ magnitude stems from the requirement of an underlying counterpart sufficiently bright at both $J$ and $H$-band and leads to a significant fraction of optically-faint SMGs not being considered here. Such sources are potentially SMGs at lower stellar mass (which will be discussed below) or the cause of overall heavy dust attenuation frequently observed in SMGs \citep[see e.g.,\,][]{Simpson2017}. Our analysis might therefore not be able to include SMGs with extreme $A_{\rm V}$ gradients towards their centers, which represents an important caveat in our analysis and will be discussed later in Section\ \ref{Results.sec}.

\begin {figure}[tb]
\centering
 \includegraphics[width=0.49\textwidth]{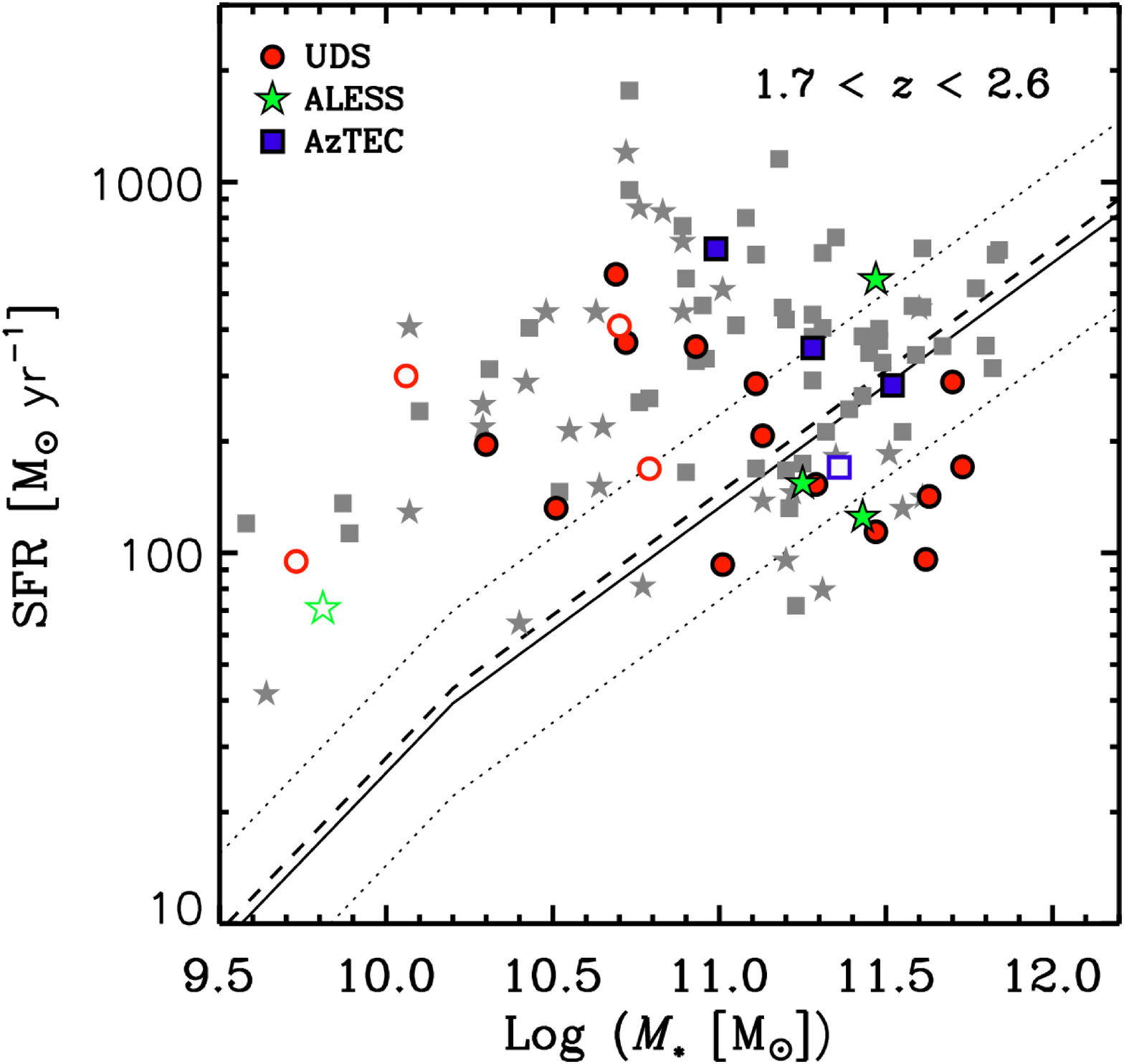}
\caption[s]{Our SMG sample shown in the $M_*$-SFR plane. Both quantities are based on SED-derived values from {\sc Magphys}. The final sample is shown as filled symbols, and SMGs rejected due to criterion 3 (i.e.,\,faint in $J$/$H$-band) are shown as open symbols. Gray symbols show the underlying parent samples of ALESS and AzTEC SMGs at $1.7 < z < 2.6$. The solid and dotted lines represent the main sequence \citep[adopted from\,][]{Whitaker2014} at the median redshift of our sample ($z= 2.15$), and the corresponding scatter ($ 0.3 \, {\rm dex}$), respectively. The dashed line denotes the median main sequence offset of 0.04 dex of our final SMG sample.}
\label{MS.fig}
 \vspace{3mm}
\end {figure}

We furthermore examine the location of our SMG sample in the $M_*$-${\rm SFR}$ plane (see Figure\ \ref{MS.fig}). Stellar masses and star-formation rates in this Figure are based on SED-derived values from {\sc Magphys}. Faint $J$/$H$-band sources rejected due to criterion 3 are shown as open symbols, for reference. As an indication for a representative parent sample of SMGs, we additionally plot the position of ALESS and AzTEC SMGs at $1.7 < z < 2.6$, as those have publicly available estimates for stellar mass and star-formation rate. We caution that the derived stellar masses for our sources might be subject to systematic uncertainties stemming from assumptions in the SED fitting such as star formation history, IMF, metallicity, and details of the dust-extinction recipe, that exceed the statistical uncertainties as output by {\sc Magphys} quoted in Table\ \ref{Sample.tbl}. More specifically, the impact of changing the assumed form of the SFH (i.e.,\,instantaneous burst versus constant model) has been demonstrated to be as large as $\sim \times 3$ in total stellar-mass estimates within SED fitting \citep{Hainline2011,Simpson2014,Michalowski2014}. In the remainder of this work, we will therefore consider these uncertainties as caveats when discussing our results. 

Our final sample overlaps well with the underlying SMG population at high stellar mass in the range $\log{(M_*/{\rm M_{\sun}})} = 10.3$-11.7. The median stellar mass of our sample, $\log{(M_*/{\rm M_{\sun}})} = 11.3$ is consistent with that of the underlying SMG population ($\log{(M_*/{\rm M_{\sun}})} = 11.1$). The star-formation rate distribution of our sample has a range of $93 < {\rm SFR} \,[{\rm M_{\sun} \, yr^{-1}}] < 661$, with a median of ${\rm SFR = 202 \pm 36 \, {\rm M_{\sun} \, yr^{-1}}}$. The median star-formation rate of the underlying SMG population is slightly higher ({\rm SFR = $326 \, {\rm M_{\sun} \, yr^{-1}}$}). Inspecting the locus of faint $J$/$H$-band sources, we find that, most notably, those reach to lower masses ($\log{(M_*/{\rm M_{\sun}})} < 10.3$). This seems plausible as those sources are rejected due to their faint optical counterparts.

The position of the star-forming main sequence at the median redshift probed by our sample is shown in Figure\ \ref{MS.fig}. The majority of our final SMGs (13 sources) lie within the scatter (i.e.,\,0.3\,dex) or below the main sequence and thus our sample shows a considerable overlap with the general star-forming population at $z\simeq 2$. This is an agreement with studies demonstrating that the massive ($\log{(M_*/{\rm M_{\sun}})} \gtrsim 11$) SMG population selected at (sub-)mm wavelengths at $z\simeq 2$ overlaps substantially with the main sequence \citep[e.g.,\,][]{Michalowski2012,Michalowski2017,daCunha2015}, which in turn is derived from galaxy populations based on other selections also including UV/optical/NIR emission. Consequently, optical/NIR-selected galaxies are also shown to exhibit (sub-)mm emission \citep[i.e.,\,][]{Tadaki2017a}. The overlap of our SMGs with the main-sequence population is reflected by an overall mild positive offset of specific star-formation rate compared to the main sequence ($\log{( {\rm sSFR/sSFR_{MS})}=0.04} \pm 0.08$). However, seven of our sources probe the regime of higher star-formation rate reaching up to {$\log{({\rm sSFR/sSFR_{MS})}=0.83}$. Systems that lie in this region of the $M_*$-SFR plane (i.e.,\,with sSFR enhancements of factors 2--3 relative to the main sequence) are commonly referred to as `starburst' galaxies \citep[e.g.,\,][]{Elbaz2011,Schreiber2015}. With our sample spanning 1.5 orders of magnitude in sSFR, it is well suited for exploring and comparing galaxy properties of starburst galaxies versus more moderate main-sequence galaxies.

\section{Methodology}

We now outline our method for deriving spatially resolved stellar mass distributions inferred from $J_{125} - H_{160}$ color maps, and obtaining structural measurements from those and the observed ALMA images. The filter combination of $J_{125}$ and $H_{160}$ probes rest-frame optical wavelengths at $z\simeq2$ and can therefore be used to infer stellar \ML\ ratios \citep[e.g.,\,][]{Bell2001}. Although imaging from additional bands are offered by the {\it HST}/CANDELS survey, we only use the $J_{125}$ and $H_{160}$ filters in this work. The reason is that the emission in those two bands is effectively detected within our sources, which is not the case for bands at shorter wavelengths.

First, we explain in Section\ \ref{color.ref} how the resolved color maps are computed. Their conversion into maps of observed mass-to-light ratio and subsequent stellar mass distributions are presented in Section\ \ref{conversion.sec} and Section\ \ref{final_maps.sec}. Finally, we describe the extraction of structural parameters from the stellar-mass maps as well as ALMA images in Section\ \ref{structural.sec}.

\subsection{Resolved stellar mass distributions}

\subsubsection{Derivation of color maps}
\label{color.ref}
We create {\it HST} $J_{125} - H_{160}$ color maps for each SMG in our sample. First, we match the PSFs of both filter bands. We construct a median-stacked PSF for each CANDELS field and filter on the basis of 5-7 well-exposed and non-saturated stars. The spatial resolution in $J_{125}$ and $H_{160}$ band based on our stacked PSFs is $0\farcs18$ and $0\farcs19$, respectively. With these PSFs, we use the PYRAF task {\sc PSFMATCH} to construct a smoothing kernel to convolve the $J_{125}$ image to $H_{160}$-band resolution, separately for each field. We test the quality of the PSF-matching by ensuring that no artificial radial color-gradients are introduced when dividing the median matched PSFs in both filters.

To ensure that all pixels in our final color maps have sufficient \SN\ in the outer regions, we perform a Voronoi-binning scheme \citep{Cappellari2003}. Within this technique, adjacent pixels are grouped together within bins that fulfill a minimum desired \SN\ threshold (named the `target \SN'). This increases the quality of resolved color distributions \citep[e.g.,\,][]{Wuyts2012,Tacchella2015b}. Before the binning is applied, all pixels in both $J_{125}$ and $H_{160}$ images that can be associated with the target SMGs are identified by creating segmentation maps with {\sc Sextractor} \citep{Bertin1996}. In cases where sources have close neighbors, we identify the main SMG component as the one associated with the submm ALMA emission. Pixels associated with emission from close neighboring sources are masked. The \SN\ thresholds for our segmentation maps are based on the average \SN\ of both $J_{125}$ and $H_{160}$ filters, and are adapted for each target individually (within the range of 1.5-3 times the background r.m.s noise). For SMGs with close neighboring galaxies, we carefully adjust the {\sc Sextractor} parameters such that the final segmentation maps only include the main SMG component. The Voronoi-binning is then performed on all pixels within the segmentation map of each galaxy, adopting a target \SN\ in the range of 10-15. We choose this range of target \SN\ to achieve an effective suppression of noise in the color distributions, especially for the outer Voronoi bins, while still keeping the bins small enough to detected radial color variations. Similar target \SN\ ratios have been used in the literature to analyze color distributions of $z\simeq2$ galaxies based on comparable {\it HST} data sets \cite[see e.g.,\,][]{Tacchella2015b}. We derive the final color maps by dividing the binned $J_{125}$ and $H_{160}$-band images. The Voronoi-binned color maps are shown in Figure\ \ref{Cutouts.fig} with a fixed color scaling for the entire sample. 

\begin {figure}[tb]
\centering
 \includegraphics[width=0.48\textwidth]{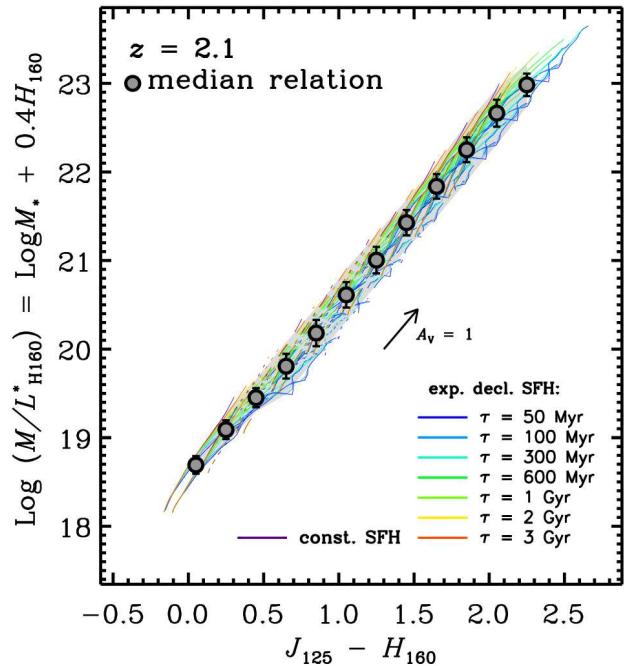}
\caption[s]{Relation between observed \MLh\ and $J_{125}$ - $H_{160}$ color. Age-tracks derived from \cite{Bruzual2003} models for a variety of SFHs shown in different colors. The ages range from 20 Myr to 3.1\,Gyr at $z = 2.1$. For each SFH, two tracks are shown that represent the range of assumed metallicities (i.e.,\,0.2 and 1 times the solar value). Different line-styles demonstrate the effect of extinction, implemented by following the \cite{Calzetti2000} description, ranging from $A_{\rm V} = 0$ (solid line) to $A_{\rm V} = 8$. Gray symbols show the median relation binned in $J_{125}$ - $H_{160}$ color. Error bars denote the 1-$\sigma$ scatter of all tracks at a given color, and the shaded polygon indicates the maximum range in \MLh\ of all models considered. There is a well defined relation that allows a robust determination of \lMLh\ at a given observed $J_{125}$ - $H_{160}$ color given our model assumptions.}
\label{ML.fig}
 \vspace{3mm}
\end {figure}

\subsubsection{Conversion from light to stellar mass}
\label{conversion.sec}

Next, we convert our $J_{125} - H_{160}$ color distributions into maps of observed stellar mass-to-light ratio in the $H_{160}$-band (i.e.,\,\MLh). The two filters probe the spectrum close to the age-and extinction sensitive Balmer break at 4000\,\AA{} rest-frame wavelength at these at $z\sim 2$ providing a robust relation between $J_{125} - H_{160}$ color and \MLh\ for our SMGs. Although additional {\it HST}/WFC3 filters at shorter wavelengths are available within CANDELS, only the $J_{125}$- and $H_{160}$-band emission (at the longest wavelengths available) effectively cover the galaxy with sufficient \SN\ for our sample. This is not the case for available bands at shorter wavelengths, for which the emission of our sources appears to be strongly or entirely suppressed in most of our targets. Therefore, we regard the information based on those filters as negligible and only analyze the $J_{125} - H_{160}$ color. As we will show below, \citep[and as shown by previous studies, e.g.,\,][]{Bell2001,Tacchella2015a}, this rest-frame optical color already yields a robust proxy for the \MLh\ quantity given the model assumptions outlined below.

We calibrate the $(J_{125} - H_{160})$-\MLh\ relation by constructing synthetic galaxy SEDs based on \cite{Bruzual2003} models. We adopt models with a \cite{Chabrier2003} IMF, two values for metallicities (0.2 and 1 times the solar value), and a variety of star formation histories, from exponentially declining (with $e$-folding timescales ranging from 50 Myr to 3\,Gyr) to constant star-formation rates. As a further ingredient, we consider a \cite{Calzetti2000} extinction law, assuming a foreground dust screen with a range of $A_V$ from 0 to 8.

Figure\ \ref{ML.fig} shows the age-tracks (ranging from 20 Myr to the age of the Universe at the given redshift) of our modeled galaxy SEDs in the $(J_{125} - H_{160})$-\lMLh\ plane for the full range of assumed SFHs, metallicities, extinctions for a redshift of 2.1. The quantity \lMLh\ is derived as the logarithmic ratio between the underlying true stellar mass and the observed light in $H_{160}$, and is shown as parametrized as ($\log M_* + 0.4 H_{160}$). To obtain the age tracks, we redshift our model SEDs and compute the observed flux in the WFC3 F125W and F160W filter. We then combine all age-tracks and compute the median $(J_{125} - H_{160})$ versus \lMLh\ relation at a given redshift, shown as filled circles in Figure\ \ref{ML.fig}. The change of this relation with redshift is shown in Figure\ \ref{ML_polygon.fig}.

All model tracks move along a well defined location within the parameter space. The main effect of the assumed foreground extinction at fixed age is to shift the relation along the age tracks, while changing metallicity hardly affects the relation. Thus, despite these degeneracies, this relation allows one to constrain \MLh\ and therefore the underlying stellar mass without prior knowledge of details of the SFHs, metallicities, or extinction at a given redshift. 
The error bars in Figure\ \ref{ML.fig} represent the 1-$\sigma$ scatter in \lMLh\ of all models considered at a given color. These errors range from $0.1$ dex at $z=1.7$ to $0.3$ dex at $z=2.6$, considering the typical range in $J_{125} - H_{160}$ color for our sample galaxies. We note that these error estimates in \lMLh\ depend on the choice of the set of model tracks used, and therefore we show the maximum range in \lMLh\ of our models at given color as shaded polygon (ranging up to $\sim 0.4$ dex at $z=2.1$). This is discussed in more detail in Appendix \ref{ML_uncertainties.sec}.

We note that deviations from a smooth star-formation history within the regions of our sources might introduce further systematic uncertainties in the above relation. \cite{Tacchella2015a} have shown that, e.g., delayed or increasing tau models occupy the same locus of parameters in the $(J_{125} - H_{160})$-\lMLh\ plane. Furthermore, episodes of star formation in addition to the smooth star-formation history might also impact our derived relation. Therefore, in Appendix\ \ref{Burst.sec}, we explore a grid of models including past burst events. In short, we find that moderate past bursts lead to a systematic increase of \lMLh\ that falls within the uncertainties. Only for the extreme cases where most of the galaxy mass was formed at redshifts $\gtrsim 3$ and where the presently ongoing star formation dominates the light, we do find that the inferred \lMLh\ is underestimated more significantly.

\subsubsection{Final stellar-mass maps}
\label{final_maps.sec}
We apply the $(J_{125} - H_{160})$-\MLh\ relation at the redshift of each SMG to obtain inferred \ML\ maps. The \ML\ maps contain only pixels that fall within the segmentation map, i.e.,\,that posses enough signal-to-noise in both filters to derive a robust $(J_{125} - H_{160})$ color estimate. Thus, we extrapolate the measured color into the faint regions dominated by background noise. This step is necessary as performing fits with {\sc GALFIT} (see Section\ \ref{structural.sec}) requires sufficient `empty' background area around the sources. We perform this step by computing the mean \MLh\ of the closest three Voronoi-bins at each pixel position around our SMGs. We then multiply these extrapolated \MLh\ maps with the $H_{160}$-band image to obtain the final stellar-mass distributions.

We compare the total stellar mass from the best-fit model to the stellar mass from {\sc Magphys} in Figure\ \ref{Mass_comp.fig}. The sum of our stellar mass maps is derived by integrating the cumulative mass profile from the best-fit mass models (as explained in Section\ \ref{structural.sec}) in large apertures ($\gtrsim 2^{\prime\prime}$) to capture the entire galaxy mass. We note that these summed stellar masses only change by $\simeq 0.02\,$ dex on average when using the actual mass maps instead of the best-fit models, or when changing the aperture to $1\farcs{5}$. The scatter of the relation in Figure\ \ref{Mass_comp.fig} is 0.3 dex. Moreover, we find a systematic offset with $M_*$(SED) being increased by about 0.3 dex with respect to our integrated stellar-mass maps. To investigate the origin of this offset, we first compute the total stellar masses of our sources derived from estimating \lMLh\ based on the galaxy-integrated $H_{160}$- and $J_{125}$-band magnitudes. We find that the masses based on integrated colors are on average lower than the summed stellar mass maps, with a median offset of $0.2\,$ dex. This is in agreement with previous studies deriving resolved mass distributions based on galaxies at low and high redshift \citep[e.g.,\,][]{Zibetti2009,Sorba2018}, and can be attributed to the luminosity weighting of galaxy color in unresolved photometry. In our case, this effect would even increase the apparent mass offset in Figure\ \ref{Mass_comp.fig}. Thus, we suggest that this offset is more likely caused by differences in the treatment of dust attenuation in {\sc Magphys} compared to the foreground dust screen assumed in our $(J_{125} - H_{160})$ to \MLh\ conversion. Furthermore, differences in the assumed shape of the star-formation histories (as pointed out in Appendix \ref{Burst.sec}) might contribute to the systematic mass offset. Other ingredients needed to build synthetic galaxy spectra, such as stellar libraries, as well as IMF for our method, are identical to the ones used by {\sc Magphys}. Moreover, the energy balance built into {\sc Magphys} might cause \MLh\ (and thus stellar mass estimates) to be elevated with respect to our $(J_{125} - H_{160})$ to \MLh\ conversion. We note that other studies in the literature found that stellar mass estimates for SMGs derived from {\sc Magphys} show similar systematic offsets (about $0.3$ dex) with respect to SED-fitting codes without an energy balance \citep[such as {\sc LePhare}; e.g.,\,][]{Gomez2018}.

In the remainder of this work, we adopt the {\sc Magphys}-derived stellar masses as the integrated mass estimates for our sample. However, we base our structural measurements on the reconstructed stellar mass maps that represent the relative mass profiles of our sources. We also address the contribution from a central `hidden' stellar mass component that we are not able to recover with our method due to our simplified assumption of foreground dust attenuation (and that might be better recovered with {\sc Magphys}) in Section\ \ref{strong_dust.sec}.

\begin {figure}[tb]
\centering
\includegraphics[width=0.49\textwidth]{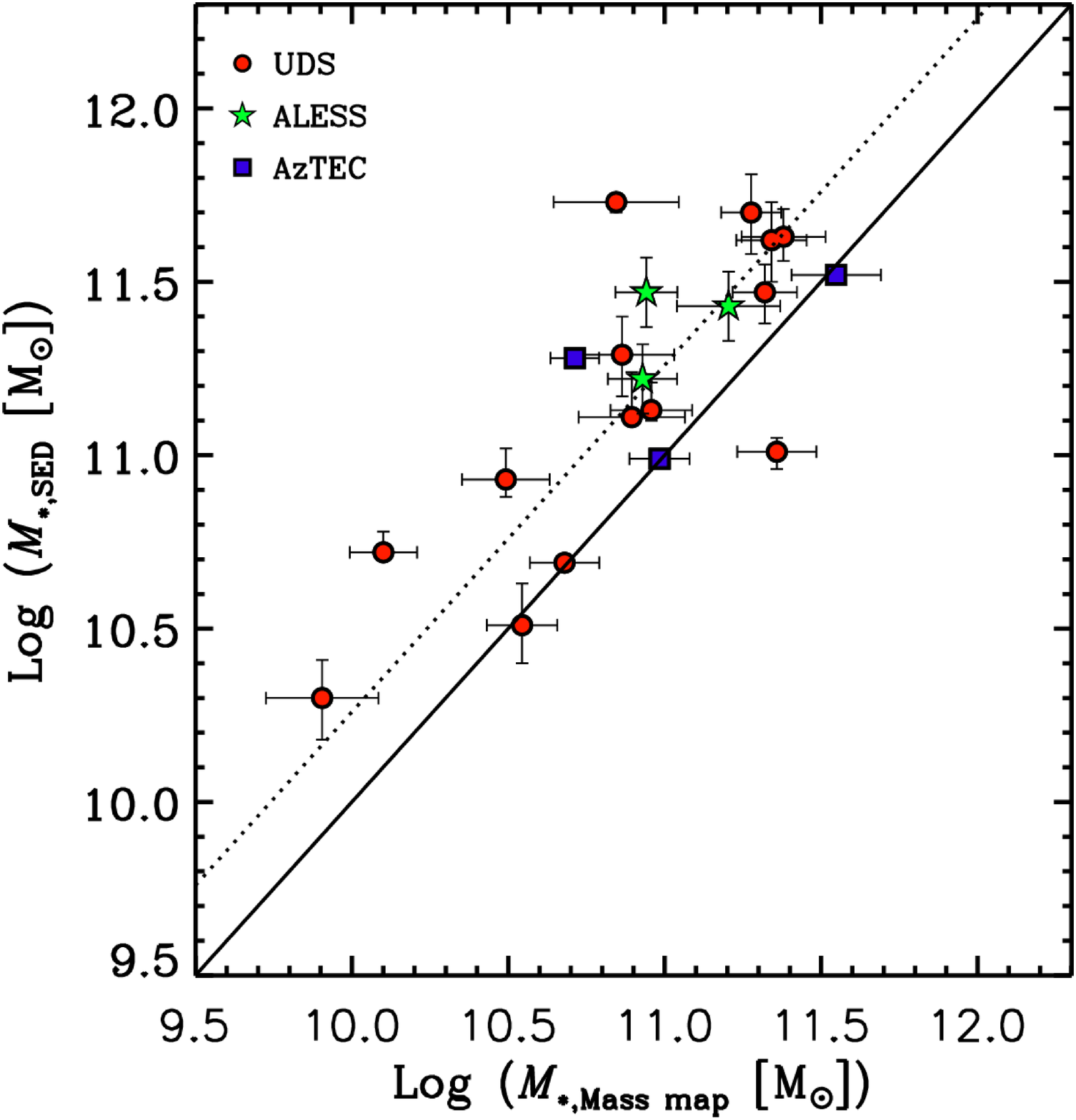}
\caption[s]{Total stellar masses derived from SED-modeling versus integrated stellar-mass maps for our SMG sample. The sub-samples taken from different SMG surveys are shown as different symbols. The solid line indicates a one-to-one relation. The integrated stellar masses from {\sc Magphys} exhibit a positive offset (0.3 dex; dotted line) with respect to the summed stellar mass maps.}
\label{Mass_comp.fig}
 \vspace{3mm}
\end {figure}

\subsection{Structural parameters}
\label{structural.sec}
To determine PSF-corrected structural parameters of our SMGs, we perform surface-brightness fits to their $H_{160}$ light distributions, stellar-mass maps, and ALMA images using {\sc Galfit} \citep{Peng2010}. {\sc Galfit} is a two-dimensional modeling code that fits parametric models to surface brightness distributions accounting for PSF convolution. We use a \Sersic\ profile as our fiducial model within the fitting, which is parametrized in terms of its intrinsic (i.e.,\,PSF-corrected) effective half-light radius along the major axis (\Re) and \Sersic\ index, $n$. Below, we briefly explain the input parameters and procedures used within {\sc Galfit} for our different data sets.

To fit both $H_{160}$ images and stellar-mass maps, we produce cutouts of $6^{\prime\prime} \times 6^{\prime\prime}$ size so that they contain sufficient sky background. We remove neighboring sources by using a mask. In the case of close galaxies with overlapping isophotes, sources are fitted simultaneously within one {\sc Galfit} run. As the input PSF for both $H_{160}$ light and stellar-mass maps, we take the stacked median PSF in the F160W filter (see Section\ \ref{structural.sec}). Error maps required by {\sc Galfit} are derived from the available CANDELS weight maps, which represent the inverse variance including various background noise terms. Then, we scale the CANDELS-derived errors such that their median value corresponds to the rms determined from the $H_{160}$ maps. When fitting the mass and light maps, we fix the galaxy center to be the mass-weighted center determined directly from the stellar-mass maps (i.e.,\,using all pixels of the stellar-mass map that are associated with the target based on the segmentation map). To estimate the accuracy of our mass-weighted centroid positions, we generate 150 versions of each mass map, each time perturbing the $J_{125}$- and $H_{160}$-band images with their associated r.m.s noise when creating our Voronoi-tessellated \MLh\ maps, and measure each time the resulting mass centroid position. 

To derive errors on the best-fit parameters of \Re\ and $n$ determined from the light and stellar mass distributions, we repeat the fitting 150 times varying the central position according to a Gaussian distribution with $\sigma$ equal to the uncertainty in the mass centroid position. The final errors of \Re\ and $n$ for a given source are then derived from the upper and lower 68\% confidence interval of the resulting distributions of \Re\ and $n$.

Similar fitting procedures are applied to our ALMA images. First, we extract thumbnails for each source and produce source masks to reject neighbors and/or fit very close-by galaxies simultaneously. As input PSFs, we create two-dimensional Gaussian images according to the major and minor beam axes and position angle as based on the clean beam. We chose to fit in the image plane since recent studies have shown that this provides consistent results compared to $uv$-fits \citep[e.g.,\,][]{Simpson2015a,Hodge2016}.
To create the error maps, we assume a constant background noise that corresponds to the background r.m.s measured directly from the ALMA images. During our fits, we allow the center position to vary freely to account for potential systematic offsets in the centroid positions between the submm and stellar components of our sources. To compute uncertainties in the best-fit parameters, as well as best-fit centroid positions measured on the ALMA maps, we repeat the above fitting process 150 times, each time perturbing the ALMA image by the r.m.s noise and computing the 68\% confidence interval of the of the resulting distributions of \Re\ and $n$. 

The uncertainty in measuring centroid positions of the stellar mass and ALMA $870\,$\mum\ components is 0.6\,kpc and 0.2\,kpc, respectively. The accuracy of measuring spatial offsets between both components is thus $\simeq 0.7$ kpc ($\sim 0\farcs1$ at $z \simeq 2$).

For presenting our following results on the comparison between the dust and stellar-mass morphologies in Sections\ \ref{dust_stellar.sec} - \ref{ETG.sec}, we only consider the subset of 14 sources covered with ALMA observations at high angular resolution (i.e.,\,${\rm FWHM} < 0\farcs4$). 

\begin{table}[t]
{\small
\centering
\caption[a]{Intrinsic effective sizes of our sample sources based on the ALMA images, $H_{160}$-band images and stellar-mass maps.}
\begin{tabular}{l*{3}{c}r}
\hline
\hline
 ID & $R_{{\rm e,submm}}$ & ${R}_{{\rm e,H160}}$ &  $R_{{\rm e, mass}}$ \\
  & [kpc] & [kpc]  & [kpc] \\
\hline 
  AS2UDS.113.1  &  $2.0^{+0.5}_{-0.4}$ &  $4.5^{+0.3}_{-0.3}$   & $1.4^{+0.2}_{-0.1}$ \\
  AS2UDS.116.0  &  $1.9^{+0.4}_{-0.3}$ &  $6.3^{+1.3}_{-3.5}$   & $1.2^{+0.4}_{-0.3}$ \\
  AS2UDS.125.0  &  $2.4^{+0.8}_{-0.6}$ &  $7.6^{+0.9}_{-0.6}$   & $4.4^{+0.4}_{-0.2}$ \\
  AS2UDS.153.0  &  $0.7^{+0.2}_{-0.1}$ &  $4.2^{+1.7}_{-0.3}$   & $5.1^{+1.3}_{-1.2}$ \\
  AS2UDS.259.0  &  $2.1^{+1.5}_{-0.6}$ &  $4.6^{+0.2}_{-0.1}$   & $3.0^{+0.2}_{-0.1}$ \\
  AS2UDS.266.0  &  $1.1^{+0.2}_{-0.1}$ &  $3.7^{+0.1}_{-0.1}$   & $3.6^{+0.1}_{-0.1}$ \\
  AS2UDS.271.0  &  $2.0^{+0.8}_{-0.8}$ &  $8.1^{+0.2}_{-0.4}$   & $3.2^{+2.0}_{-0.8}$ \\
  AS2UDS.272.0  &  $1.4^{+0.2}_{-0.2}$ &  $4.9^{+0.2}_{-0.2}$   & $3.1^{+0.3}_{-0.3}$ \\
  AS2UDS.297.0  &  $1.2^{+0.3}_{-0.2}$ &  $3.1^{+0.3}_{-0.1}$   & $2.2^{+0.5}_{-0.2}$ \\
  AS2UDS.311.0  &  $1.3^{+0.2}_{-0.2}$ &  $4.1^{+0.2}_{-0.1}$   & $2.6^{+0.2}_{-0.2}$ \\
  AS2UDS.322.0  &   -                  &  $4.0^{+0.6}_{-0.5}$   & $0.3^{+1.0}_{-0.3}$ \\
  AS2UDS.412.0  &  $1.7^{+1.2}_{-0.5}$ &  $3.4^{+0.2}_{-0.2}$   & $0.9^{+0.1}_{-0.1}$ \\
  AS2UDS.583.0  &  $2.3^{+0.7}_{-0.5}$ &  $4.9^{+0.2}_{-0.2}$   & $3.6^{+1.2}_{-0.7}$ \\
  AS2UDS.659.0  &  $2.1^{+0.2}_{-0.3}$ &  $5.2^{+0.2}_{-0.3}$   & $0.8^{+0.7}_{-0.8}$ \\
  ALESS018.1    &  -                   &  $6.2^{+0.1}_{-0.1}$   & $2.8^{+0.1}_{-0.1}$ \\
  ALESS067.1    &  $2.1^{+0.4}_{-0.3}$ &  $5.8^{+0.7}_{-1.6}$   & $2.9^{+0.3}_{-0.2}$ \\
  ALESS079.2    &  -                   &  $6.7^{+0.2}_{-0.1}$   & $0.52^{+0.02}_{-0.01}$ \\
  AzTECC33a     &  -                   &  $5.7^{+1.8}_{-1.0}$   & $1.8^{+0.6}_{-0.3}$ \\
  AzTECC38      &  -                   &  $3.9^{+0.1}_{-0.1}$   & $2.6^{+0.1}_{-0.1}$ \\
  AzTECC95      &  -                   &  $4.1^{+0.3}_{-0.2}$   & $2.6^{+0.2}_{-0.1}$ \\
\hline
\hline
\end{tabular}

\label{Sizes.tbl}}
\end{table}

\section{Results}
\label{Results.sec}

\subsection{Rest-optical versus stellar-mass morphology}
\label{morph.sec}

\begin {figure*}[tb]
\centering
   \includegraphics[width=0.85\textwidth]{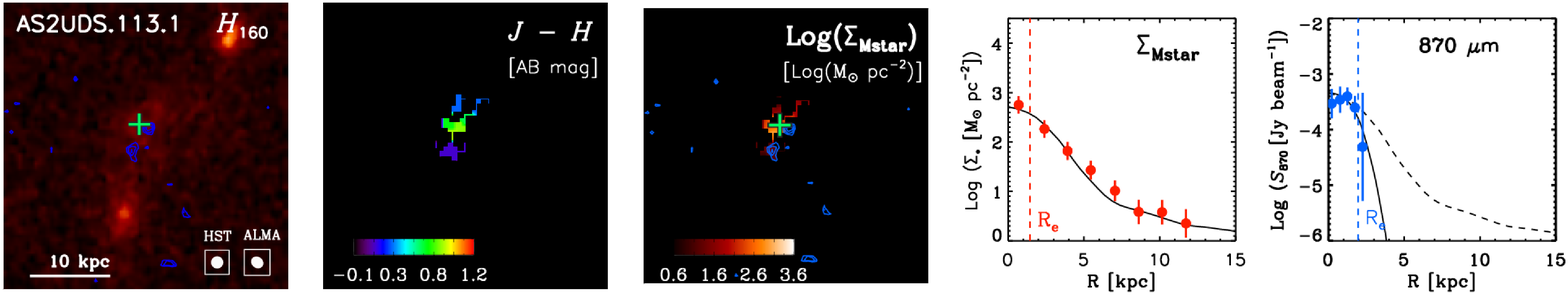}
  \includegraphics[width=0.85\textwidth]{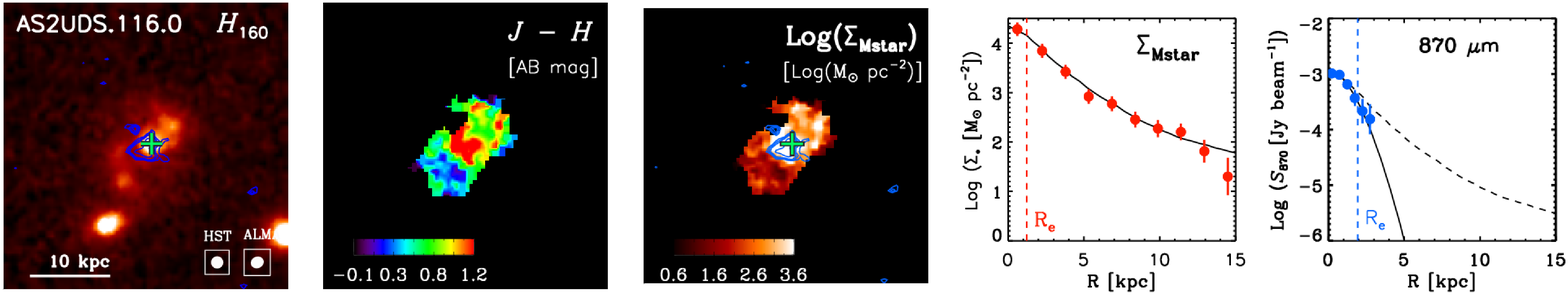}
   \includegraphics[width=0.85\textwidth]{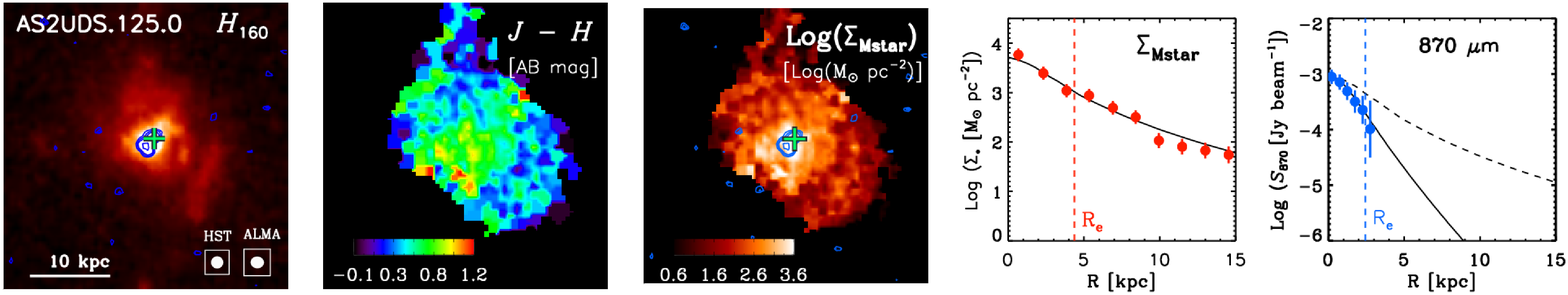}
    \includegraphics[width=0.85\textwidth]{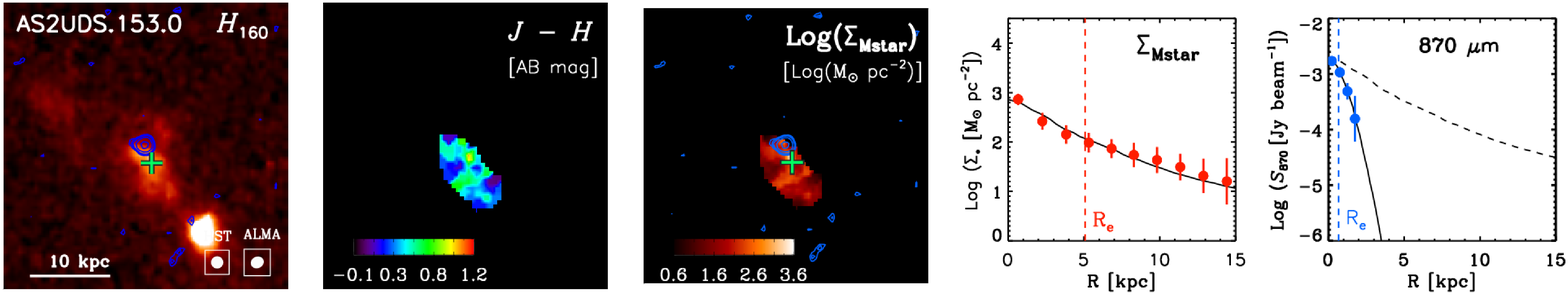}
     \includegraphics[width=0.85\textwidth]{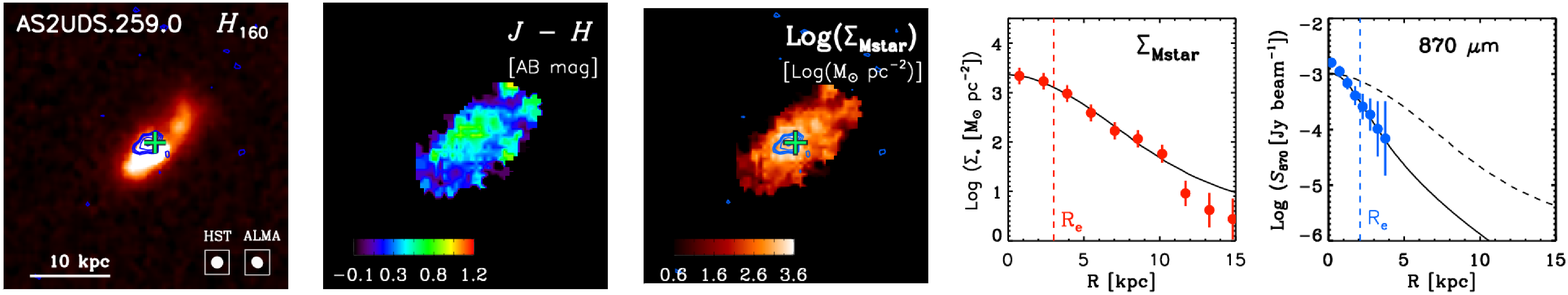}
       \includegraphics[width=0.85\textwidth]{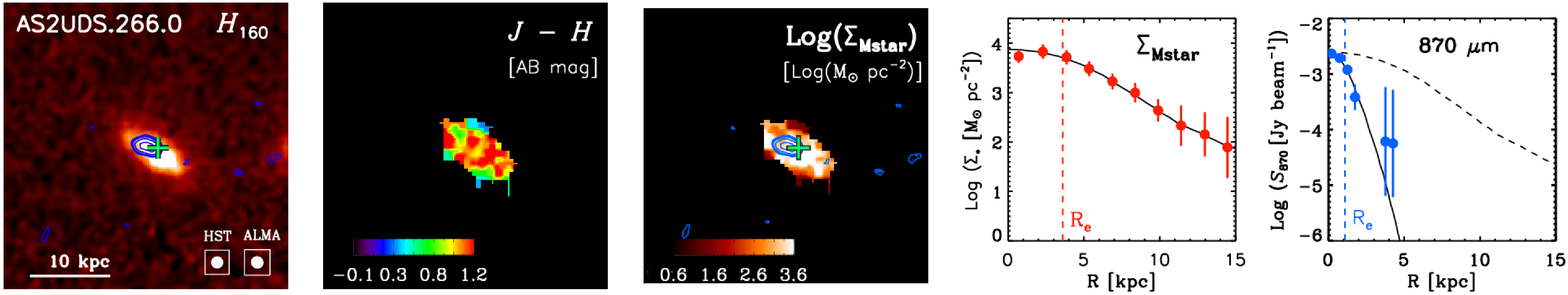}

\caption[s]{From left to right: $H_{160}$-band cutouts; $(J_{125} - H_{160})$ color maps; stellar-mass distributions (all with a fixed size of $4\farcs3$); radial stellar mass; and far-infrared ALMA profiles for our SMG sample. The scaling and dynamic ranges for each panel are kept fixed among all SMGs. For color and mass maps, only pixels associated with the main SMG (based on the segmentation map) are shown. On top of each $H_{160}$-band cutout and mass map, ALMA contours for the $S/N$ levels 2.5, 3.5, 5, 8, and 12 are shown as blue contours. The {\it HST} and ALMA PSFs are indicated as filled ellipses in the bottom right corner of the $H_{160}$-band cutouts. Mass-weighted centers are marked as crosses. Solid lines show our best-fit {\sc Galfit} models to the radial profiles. Additionally, we indicate the best-fit mass model re-normalized to the peak of the best-fit submm profile for comparison. Intrinsic effective radii are furthermore indicated as vertical dashed lines. }
\label{Cutouts.fig}
 \vspace{3mm}
\end {figure*}

\begin {figure*}[th!]
\ContinuedFloat
\centering

       \includegraphics[width=0.84\textwidth]{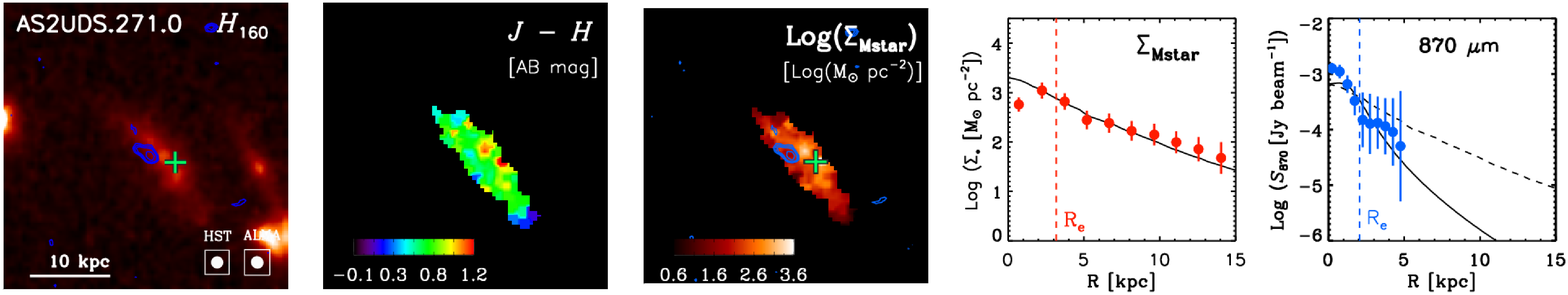}
       \includegraphics[width=0.84\textwidth]{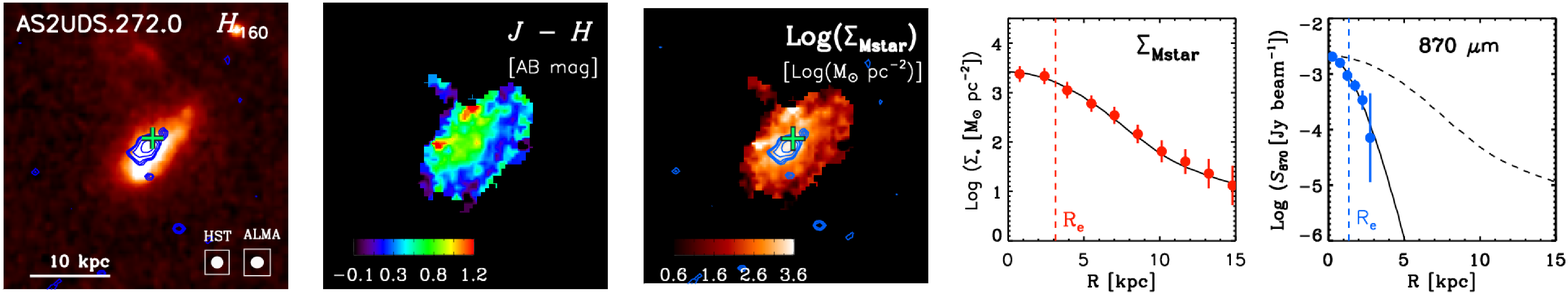}
  \includegraphics[width=0.84\textwidth]{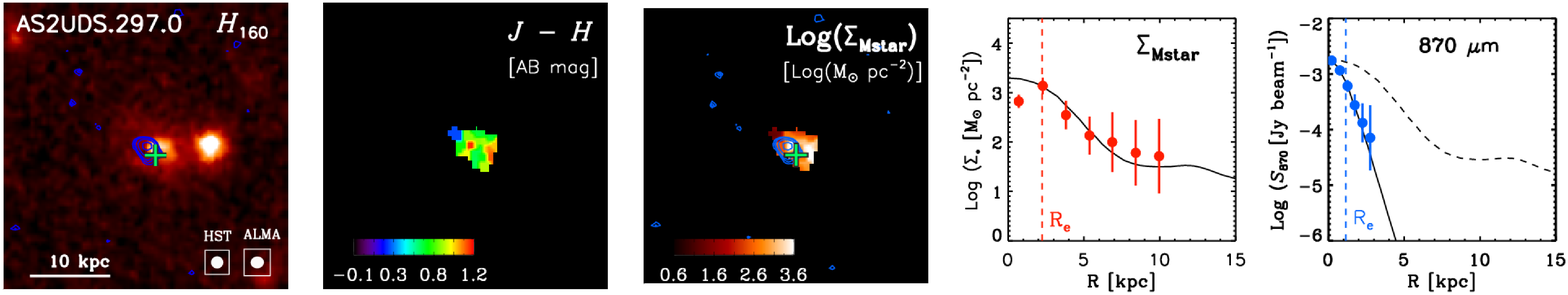}
   \includegraphics[width=0.84\textwidth]{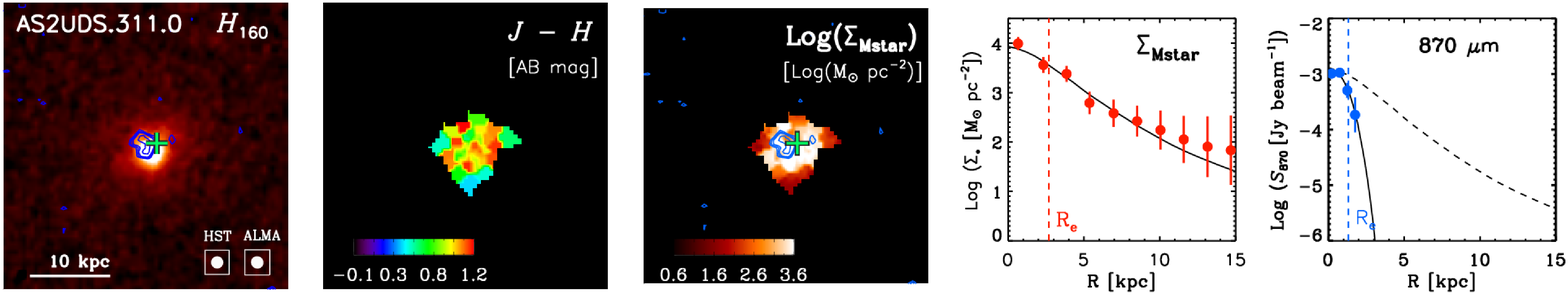}
    \includegraphics[width=0.84\textwidth]{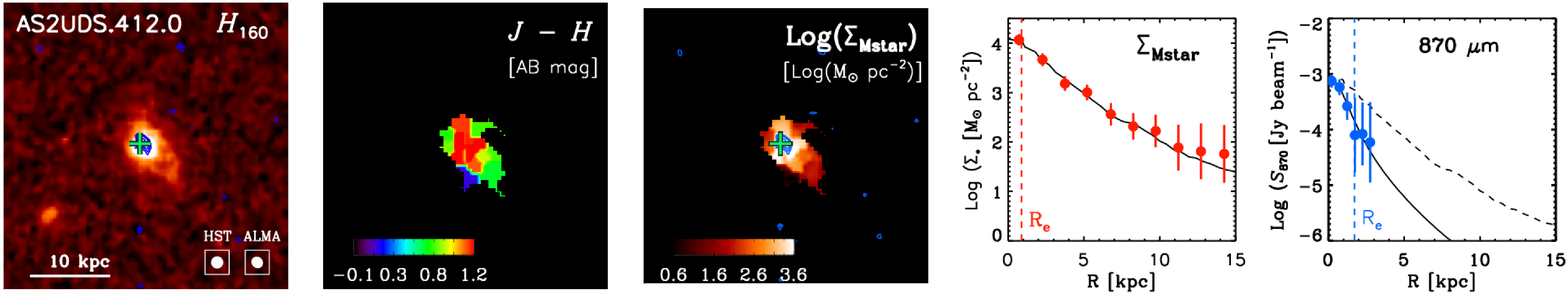}
     \includegraphics[width=0.84\textwidth]{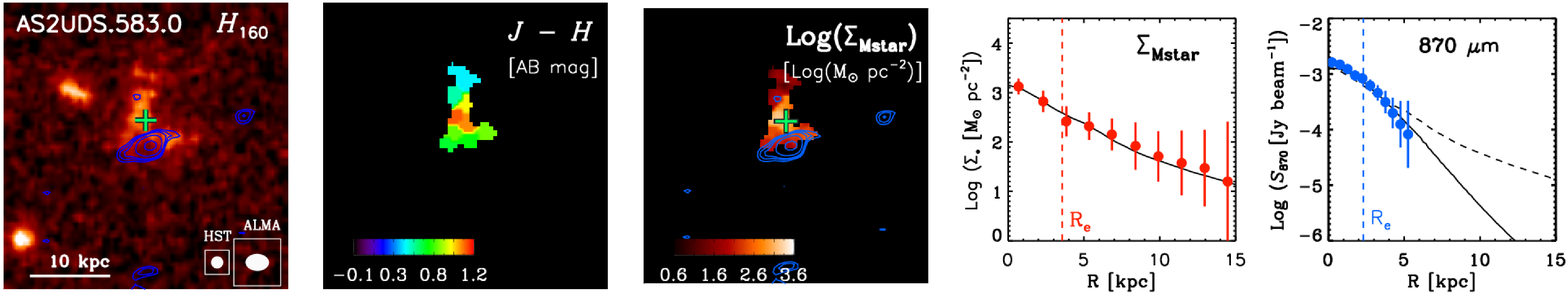}
           \includegraphics[width=0.85\textwidth]{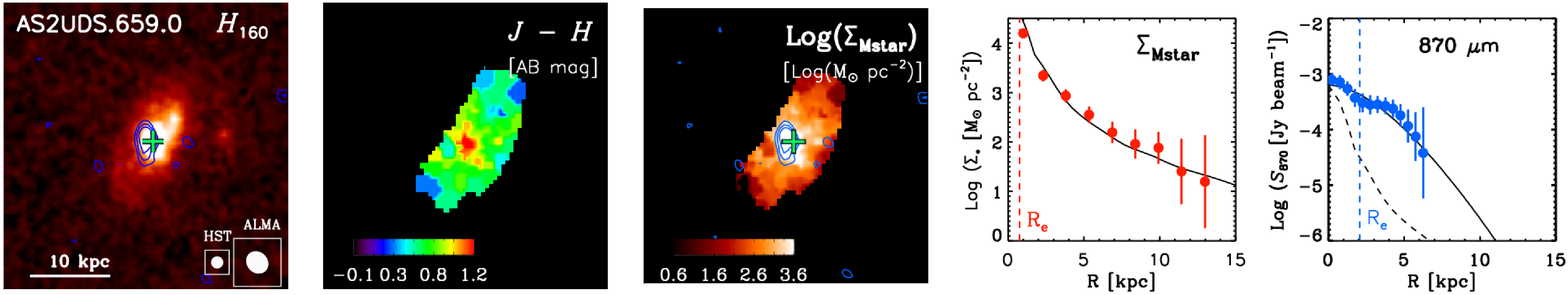}

\caption[s]{ - continued}
 \vspace{3mm}
\end {figure*}

\begin {figure*}[tbh]
\ContinuedFloat
\centering

               \includegraphics[width=0.85\textwidth]{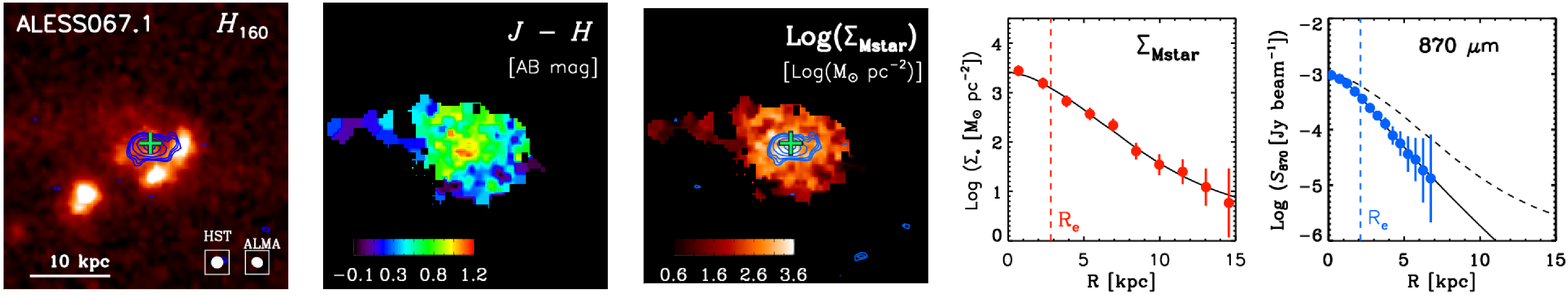}
          \includegraphics[width=0.84\textwidth]{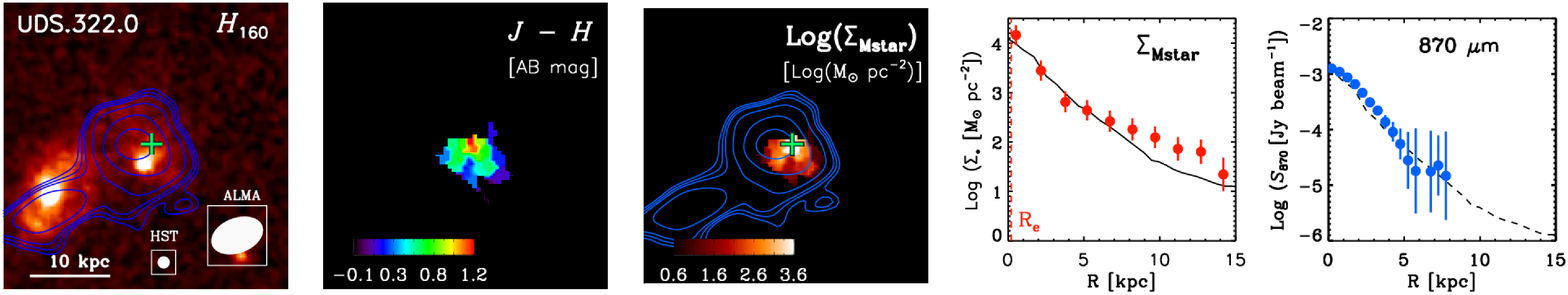}
       \includegraphics[width=0.85\textwidth]{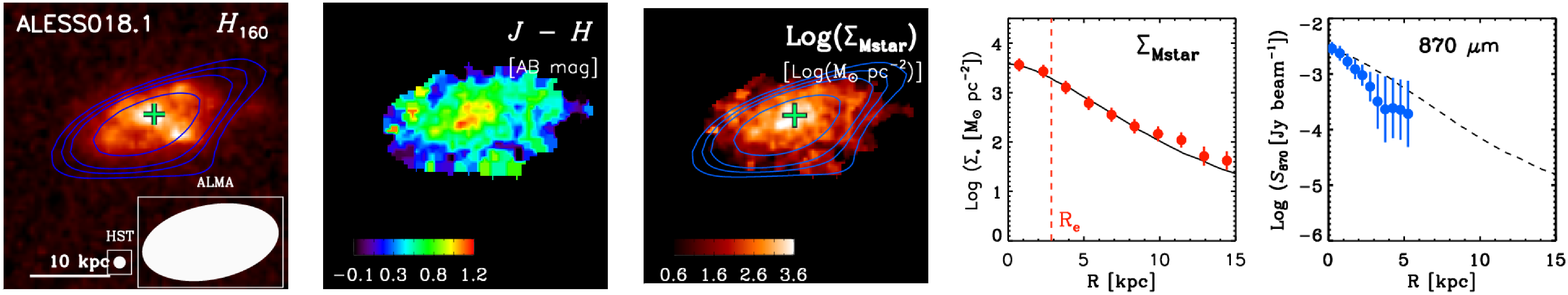}
  \includegraphics[width=0.85\textwidth]{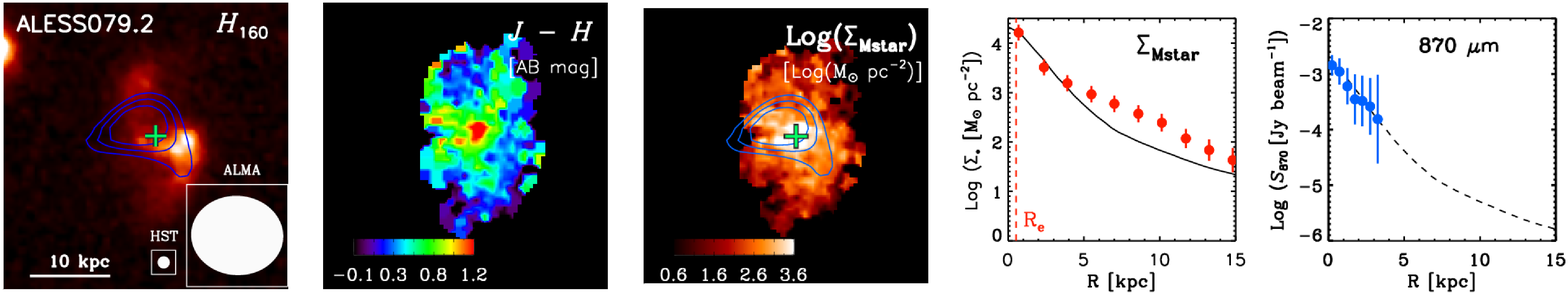}
   \includegraphics[width=0.85\textwidth]{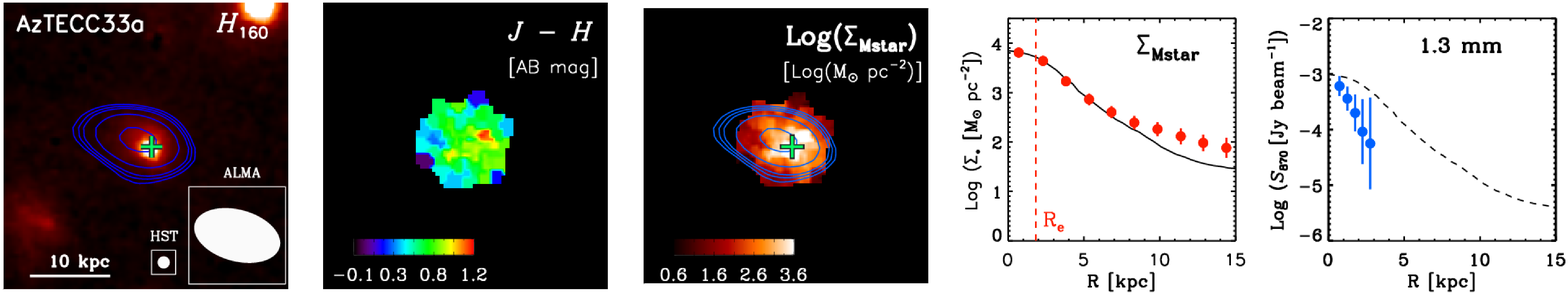}
    \includegraphics[width=0.85\textwidth]{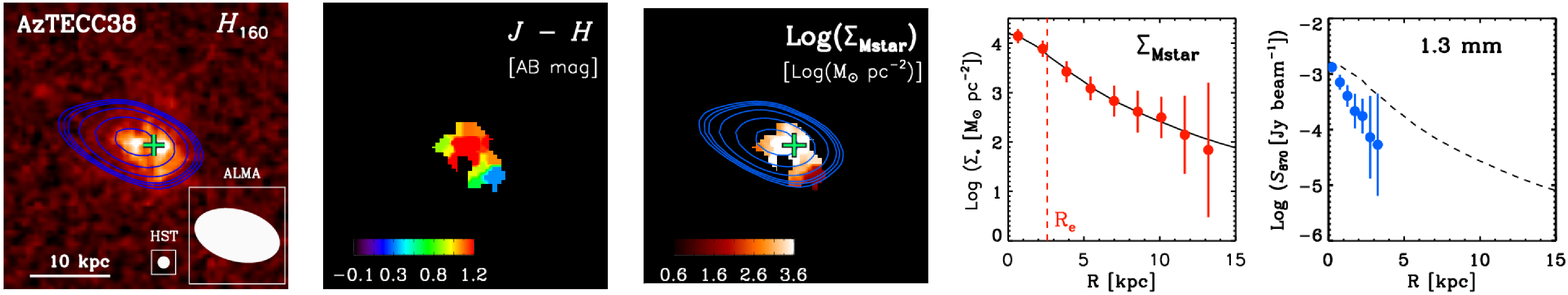}
         \includegraphics[width=0.85\textwidth]{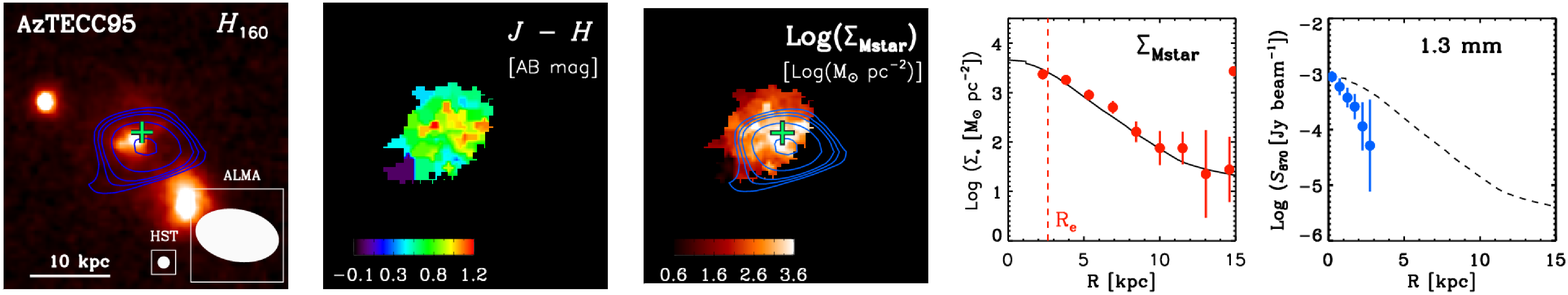}

\caption[s]{ - continued}
 \vspace{3mm}
\end {figure*}

First, we compare the resulting stellar mass distributions to the $H_{160}$-band light morphology of our SMGs sample. In Figure\ \ref{Cutouts.fig}, we show the $H_{160}$-band cutouts, $J-H$ color maps, and resulting stellar mass maps for our full SMG sample. For color and mass distributions, only regions associated with the main SMG component are shown, using a fixed spatial and intensity scaling for all targets. In case of close neighboring sources (e.g., AS2UDS.297.0), the main SMG component is defined as the source with the associated ALMA/submm emission. 

The majority of SMGs exhibits systematically redder colors towards their centers. Similarly, off-centered emission features, such as clumps or tail-like structures, appear as blue regions in the color maps. As redder colors result in higher \ML\ ratios, this implies systematic radial gradients towards higher \ML\ in the centers of our galaxies. For sources where those trends are strongest, the difference in \MLh\ along the galactocentric radius is up to 1-2 dex (in the case of e.g., AS2UDS.116.0, AS2UDS.659.0, ALESS067.1, ALESS079.2). This exceeds the systematical uncertainties of \MLh\ at a given $(J_{125} - H_{160})$ color discussed in Section\ \ref{conversion.sec}, and therefore likely reflects true spatial variations of \MLh\, even in the presence of potential variations in e.g., the star formation history as a function of galactocentric radius. In some cases, off-centered clumps dominating the light distribution in $H_{160}$-band (such as for ALESS067.1) but only weakly contribute to the stellar mass density. Moreover, the stellar mass distribution of ALESS079.2 appears as a large system with a smooth and strongly centrally peaked mass profile, rather than being comprised of several components as the $H_{160}$-band image suggests. Those cases highlight the caveat of interpreting highly disturbed rest-frame optical morphologies commonly seen in SMGs \citep[e.g.,\,][]{Swinbank2010b,Chen2015,Hodge2016}, and demonstrate that the underlying stellar mass distribution can significantly differ from the observed $H_{160}$-band distribution.

The systematic radial trends of \MLh\ within our sources are either caused by variations of stellar age and/or the effects of extinction, as those two effects cannot be distinguished on the basis of the observed $J_{125} - H_{160}$ color alone. However, in cases where the ALMA/submm emission peak coincides well with the location of strong color variations (e.g., AS2UDS.116.0, AS2UDS.659.0), the redder colors are likely the effect of increased extinction towards stronger dust-obscured regions. We note that the optical depth towards regions that are associated with strong dust-obscured star formation within SMGs might lead to high optical extinction ($A_V >> 1$) that cannot be recovered by our method. We discuss this potential caveat and its implication for our results in Section\ \ref{strong_dust.sec}. Similarly, targets with strong dust extinction gradients or overall high dust extinction, leading to a low surface brightness in the observed $J_{125}$ and/or $H_{160}$ bands might be not considered in this study due to our selection effects. We therefore emphasize that strong systematic color gradients might be even more frequent among the SMG population at high redshift compared to our sample of SMGs. Future observations from e.g., the {\it James Webb Space Telescope} ({\it JWST}) will be needed to test such conclusions. 

The observed spatial \MLh\ variations within our sources also lead to consistent changes in their radial profiles.
Table\ \ref{Sizes.tbl} lists the intrinsic effective sizes of our sample as determined on both $H_{160}$ light and stellar-mass distributions (i.e.,\,${R}_{{\rm e,H160}}$ and $R_{{\rm e,mass}}$, respectively). We derive a median effective size of our light and mass distributions of $R_{{\rm e,H160}}=4.8 \pm 0.3$ kpc and $R_{{\rm e,mass}}=2.7 \pm 0.3$ kpc, respectively. These values imply that the stellar mass components are are systematically smaller than those of the $H_{160}$-band light, with a median ratio of $\langle R_{{\rm e,mass}}/{R}_{{\rm e,H160}}\rangle = 0.5 \pm 0.1$. This is consistent with the radial \MLh\ variation observed in our sources, as higher \MLh\ at low galactocentric radii lead to overall smaller sizes. Similarly, the presence of off-centered clumpy emission in the $H_{160}$-band light as well as systematic differences in the inferred mass and light-weighted centers contribute to the inferred $H_{160}$-band sizes being more extended than the stellar mass. We will also show that this systematic size difference is independent of radial profile parametrizations (discussed in Section\ \ref{dust_stellar.sec}). 

We note that the extrapolation of color and hence \ML\ when creating our mass maps likely introduces additional systematic uncertainties in the outer stellar mass profiles shown in Figure\ \ref{Cutouts.fig}. We expect that this, in turn, most significantly affects the measured profile shape (hence \Sersic\ index). However, since the extrapolation of \ML\ only affects our mass maps outside the segmentation maps (which contain on average about 90\,\% of the total mass of our sources), this is unlikely to impact our measurements and conclusions regarding effective half-mass sizes discussed below.

The distribution of \Sersic\ indices inferred from the stellar mass has a median of $n_{{\rm mass}} = 1.4 \pm 0.4$. This compares to a median value for the light distribution of $n_{{\rm H160}} = 1.0 \pm 0.2$, hinting that the stellar mass might be slightly more centrally concentrated than the light. A few SMGs among our sample exhibit steep inner mass profiles ($n_{{\rm mass}} \ge 4$) and high central stellar mass surface densities (e.g., AS2UDS.659, ALESS79.2), suggesting that those SMGs might host centrally concentrated mass distributions. Due to the aforementioned additional systematic uncertainties in \Sersic\ index from the \ML\ extrapolation and structured central dust, we will not draw any further conclusions from the inferred $n_{{\rm mass}}$ values in this work.

Clearly, our findings highlight the importance of correcting for radial \ML\ variations when determining sizes and morphologies of the underlying stellar mass based on rest-frame optical imaging of high-redshift SMGs. 

\subsection{Dust vs. stellar-mass morphology}
\label{dust_stellar.sec}

\begin {figure}[tb]
\centering
 \includegraphics[width=0.49\textwidth]{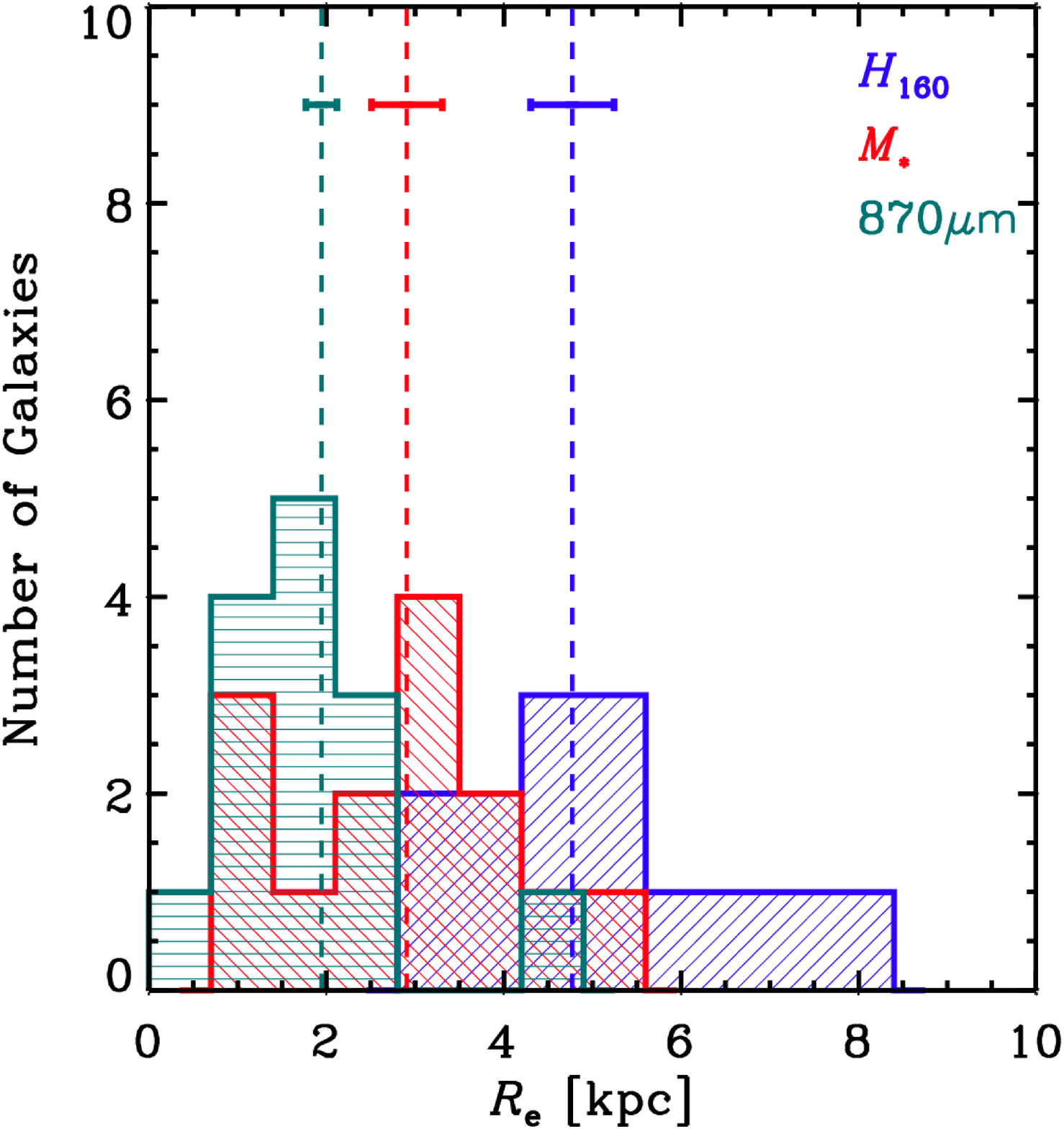}
\caption[s]{Histogram of intrinsic effective radii \Re\ as measured in $H_{160}$-band light, stellar mass, and ALMA $870\,$\mum\ emission. Median sizes are indicated as vertical dashed lines, and horizontal error bars show the respective errors on the median values. Stellar mass sizes are clearly smaller than inferred from $H$-band light, with the ALMA emission being even more compact. }
\label{Sizes.fig}
 \vspace{3mm}
\end {figure}

Next, we compare the morphology of the existing stellar mass to the dust component of our SMGs, as best approximated by our stellar mass maps and ALMA submm imaging, respectively.

The best-fit radial profiles of the 870 $\mu$m dust emission in comparison to the inferred stellar mass at same spatial resolution are directly compared in Figure\ \ref{Cutouts.fig}. These give the overall impression that the dust resides in a more compact configuration than the stellar component for the majority of our objects. To quantify this trend, we show the distribution of effective radii (\Re) derived from stellar mass maps and the ALMA submm continuum emission of our SMG sample, including the measurements on the $H_{160}$-band images, in Figure\ \ref{Sizes.fig}. Median sizes of the three different components are indicated as vertical dashed lines. In general, the ALMA emission is more compact than the stellar mass, with a median size of $\langle R_{{\rm e,submm}} \rangle = 2.0 \pm 0.1 \,{\rm kpc}$. The median intrinsic size difference between the submm emission and the stellar mass is $\langle {R}_{{\rm e,submm}}/{R}_{{\rm e,mass}} \rangle = 0.6 \pm 0.2$, demonstrating that the dust emission is more compact than the existing stellar mass for our SMG sample. For comparison, we find that this size ratio is even smaller when considering the $H_{160}$-band (with $\langle {R}_{{\rm e,submm}}/{R}_{{\rm e,H160}} \rangle = 0.34 \pm 0.03$). 

\begin {figure}[htb]
\centering
 \includegraphics[width=0.49\textwidth]{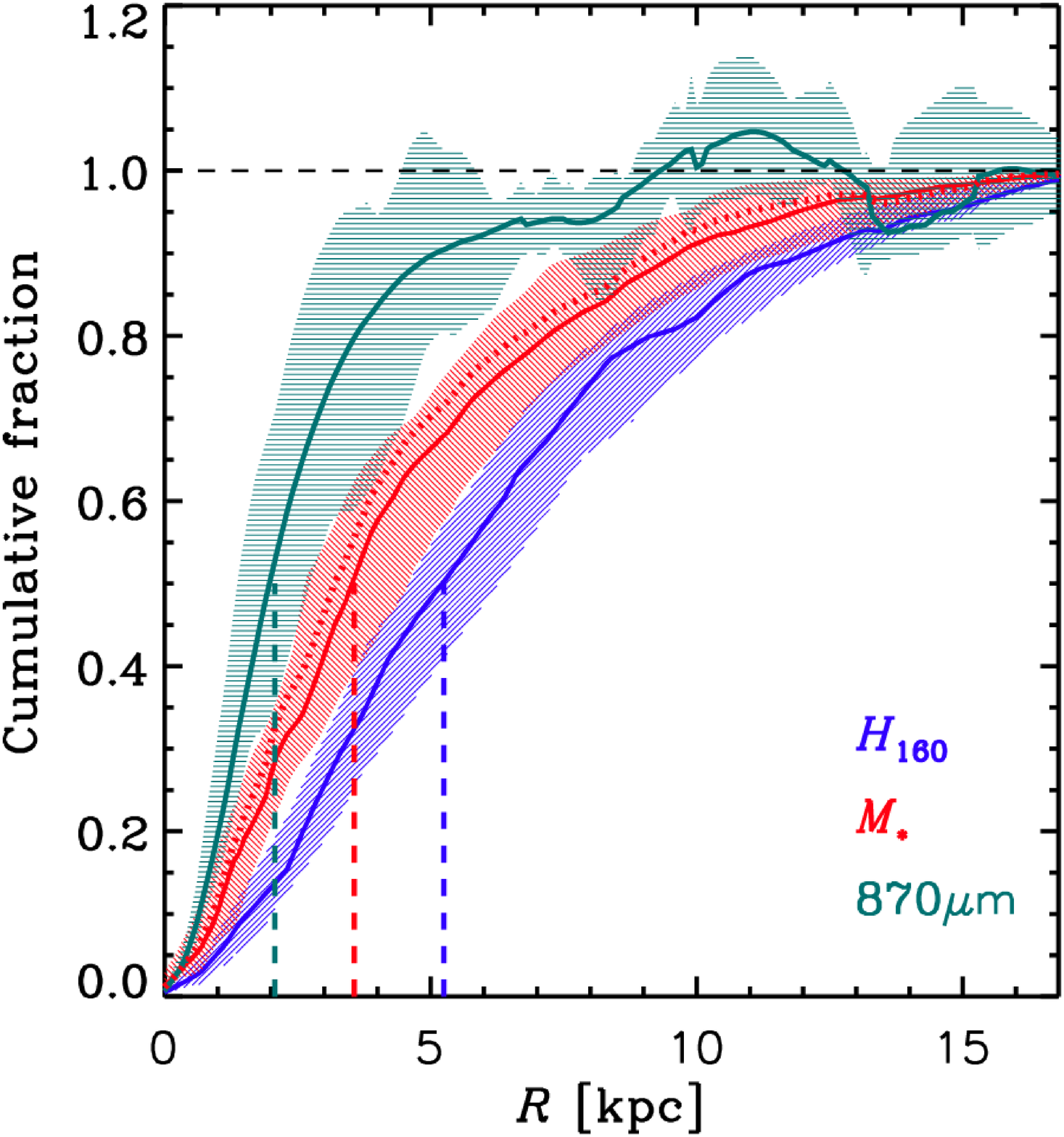}
\caption[s]{Median cumulative flux, mass and light distributions for our SMGs, as measured in $H_{160}$-band light (blue), stellar mass (red), and ALMA $870\,$\mum\ emission (cyan). Shaded areas denote the 68th percentile of all individual profiles at a given radius. Observed effective radii for all components are shown as vertical dashed lines. The red dotted line represents the stellar mass profile when considering an additional `hidden' stellar mass component derived in Section\ \ref{strong_dust.sec}.}
\label{Cumulative.fig}
 \vspace{3mm}
\end {figure}

To confirm these size differences independent of assumed radial profile shape, we consider median cumulative profiles of $H_{160}$-band light, stellar mass, and submm emission in Figure\ \ref{Cumulative.fig}. These profiles represent the median of all individual normalized profiles at a given radius, with the 68th percentile around the median being indicated by shaded areas. The individual profiles are constructed by summing up our images in elliptical apertures, and are therefore not corrected for convolution with the respective PSF. The shapes of the ellipses are computed from the central positions, axis ratios, and position angles of the best-fit {\sc Galfit} solutions for each SMG (with neighboring objects masked out). Since the ALMA images contain significant correlated background noise, we truncate the individual profiles at radii where the integrated source flux reaches a plateau (i.e.,\,indicating that the total source flux has been reached). The apparent fluctuation in the cumulative 870-$\mu$m flux level at the outer radii ($\gtrsim 10\,$kpc) arises due to these correlated noise structures. Variations of physical resolution due to the range of redshifts probed are small at fixed angular resolution ($< 6 \%$) and are thus neglected here.

These profiles confirm findings that the submm emission is significantly more compact than the $H_{160}$-band light \citep[see also\,][Gullberg et al. 2019]{Simpson2015b,Hodge2016}. Moreover, Figure\ \ref{Cumulative.fig} confirms that the stellar mass sizes are on average more compact compared to the $H_{160}$-band light, caused by the systematic radial trends in inferred \MLh. 
However, the submm emission traced by ALMA represents the most compact component probed in our sources, independent of the assumed radial profile shape. The dashed line in Figure\ \ref{Cumulative.fig}, derived from our simplified toy-model discussed in Section\ \ref{strong_dust.sec}, demonstrates that an additional stellar mass component potentially `hidden' by strong dust attenuation does not change these conclusions. 

Inspecting our resolved ALMA emission and stellar-mass maps shown in Figure\ \ref{Cutouts.fig}, the centroid position of the dust distribution agrees with a median offset of 1.1\,kpc to that of the stellar distribution. In view that the accuracy of determining this spatial separation is 0.7\,kpc (see Section\ \ref{structural.sec}), our measurement indicates that there are small intrinsic spatial displacements between the dust and stellar components. However, there is only a minor subset of three targets within our sample, for which the dust emission is clearly offset $\ge 2.5\,{\rm kpc}$ (AS2UDS.153.0, AS2UDS.271.0 and AS2UDS.583.0). The overall good spatial agreement between the dust and stellar components of our SMG sample represents an important new addition to previous studies that reported apparent spatial offsets between the optical light and submm emission \citep[e.g.,\,][]{Chen2015,Hodge2016}. Interestingly, this is also the case for some galaxies in our sample (e.g., ALESS067.1), where there is a clear offset seen between the peak of the $H_{160}$-band and submm emission. However, the spatial offset decreases when considering the centroid position in stellar mass. Overall, the centroid positions of the dust and $H_{160}$ light emission show on average larger offsets ($1.5\,{\rm kpc}$) than inferred from stellar mass.

\subsection{The case of strongly dust-obscured centers}
\label{strong_dust.sec}  

We note that our simplified assumptions on the dust geometry (i.e.,\,foreground screen) used for our models are challenged by observations of SMGs, using far-infrared and submm tracers of dust emission. In particular, high-redshift SMGs are found to exhibit high column densities of dust, implying values of $A_{\rm V} \sim 500$ (assuming that the dust in uniformly distributed within the effective submm size), even ranging up to $A_{\rm V}\sim 2000$ for the most extreme cases \citep[e.g.,\,][]{Simpson2017,Gomez2018}. Similarly, the distribution of dust does not seem to necessarily manifest itself in the attenuation of UV and/or optical emission which is detectable at the same spatial location, as has also been inferred from deviations of infrared-luminous galaxies in the IRX-$\beta$ plane \citep[e.g.,\,][for a review]{Casey2014,Popping2017}. All this implies that our simplified corrections likely fails to entirely recover the underlying stellar mass, which might partly be hidden in strongly dust-obscured regions responsible for most of the submm emission. Therefore, our above conversion from light to mass distributions represents a lower limit to the true radial \ML\ gradients discussed in Section\ \ref{morph.sec}.

In order to test the impact of an additional `hidden' stellar mass component on our averaged stellar-mass profile, we perform the following tests. First, we compute the stellar mass of such a potentially hidden component that might be produced in the recent burst event. We assume a burst timescale of $150$ Myr, although note that this gives a conservative estimate since typical burst timescales of SMGs are estimated to be around $100$ Myr \citep[e.g.,\,][]{Simpson2014}. Considering the median star-formation rate, a hidden burst component could produce on average $ 20 \, \%$ of the total stellar mass of our sample considered in Figure\ \ref{Cumulative.fig}. We add this burst mass to the total cumulative stellar mass profile by distributing the burst mass according to the averaged cumulative dust profile, assuming that the central star-forming component is well traced by the submm emission. We then re-normalize the resulting cumulative stellar mass profile. The resulting profile, shown as the dotted red line in Figure\ \ref{Cumulative.fig}), is steeper than our median stellar profile. However, the difference to our median stellar-mass profile is only marginal compared to the difference between the dust and stellar mass profiles of our sample. Hence, we do not expect the effect of additional stellar mass hidden behind a strongly dust-enshrouded central starburst to affect the results and conclusions made in this work. 

We note that our assumptions for this exercise are simplified. More specifically, we expect a hidden stellar mass component to be even more compact in case the central submm emission is optically thick and thus fails to well trace the central star-forming component. Moreover, additional past burst events and potentially evolved stellar populations in the center of SMGs might further increase the amount of hidden stellar mass not considered here. These would result in an even more compact configuration of stellar mass (i.e.,\,such that the stellar mass is closer in size to the dust distribution) than implied by our test considered here.

\subsection{Systematic trends with global SF properties}
\label{size_ratios.sec}
Next, we relate the relative radial distributions of the submm and stellar components of the SMGs in our sample to their global star formation properties, to investigate the connection between their structure and evolutionary state.

First, we quantify the dust versus stellar morphology by the relative size ratio between the submm and stellar-mass components (\rr). In Figure\ \ref{Relations.fig}, we show the relation of this quantity versus specific star-formation rate. The specific star-formation rate and corresponding offset relative to the main sequence is commonly used to distinguish the overall population of `normal' star-forming galaxies from galaxies in starburst mode that systematically lie above the main sequence \citep[e.g.,\,][]{Elbaz2011,Schreiber2015}. Typical thresholds for selecting starburst galaxies have factors of 2-3 enhancement of specific star-formation rate relative to the main sequence (corresponding to $\log{({\rm sSFR/sSFR_{MS}})} \ge 0.5)$. Our shown set of SMGs cover about 1.5 orders of magnitude in specific star-formation rate, sampling the locus of main sequence galaxies (shown as shaded area in Figure\ \ref{Relations.fig}) up to the starburst regime with a sufficient offset in specific star-formation rate to the main sequence to explore potential systematic differences among these two galaxy populations. 

The typical error margins in \rr\ and specific star-formation rate are indicated in Figure\ \ref{Relations.fig}. The uncertainties in specific star-formation rate are estimated considering typical uncertainties in \MLh\ based on our mass-to-light conversion as derived in Section\ \ref{conversion.sec}, as well as typical errors in star-formation rate based on the {\sc Magphys} output. We stress that the resulting average error margin in specific star-formation rate of our sample might be underestimated, given the potentially significant systematic uncertainties in stellar mass for SMGs (as discussed in Section\ \ref{final_sample_prop.sec}). Furthermore, all our sources shown are color-coded by their total stellar mass based on {\sc Magphys}.

We find no significant correlation between specific star-formation rate and the dust versus stellar morphologies of our sample. Instead, our SMGs exhibit compact dust versus extended stellar morphologies (with \rr\ $\le 0.5$) within the entire range of specific star-formation rate explored. Inspecting the dependency on total stellar mass, we find no correlation with \rr\ at fixed sSFR, and likewise we find no correlation of \rr\ with sSFR at fixed total stellar mass. Due to the position of our sample in the $M_*$-SFR plane (see Figure\ \ref{MS.fig}), galaxies with higher sSFR are more massive, leaving us with limited range of sSFR spanned at fixed stellar mass. Due to our sample design, we are thus not able to further explore possible correlations of \rr\ and sSFR for a given stellar mass that potentially arise within a wider dynamic range of sSFR.

We therefore conclude that compact dust cores embedded in a more extended stellar configuration found in previous studies of SMGs \citep[e.g.,\,][]{Chen2015,Hodge2016} are not confined to systems classified as starbursts based on their elevated specific star-formation rate values with respect to the main sequence. Previous studies exploring the dust emission in main sequence galaxies at similar redshift and stellar mass as explored here ($2.2 < z < 2.5$, $\log{(M_*/{\rm M_{\sun}})} > 11$), but selected to be extended rotating disks based on their ionized gas kinematics, could demonstrate that their average sample exhibits compact dust cores \citep{Tadaki2017a,Tadaki2017b}, confirming this conclusion.

In addition to galaxies with compact dust cores, some sources within our sample show submm components being more extended than the stars (AS2UDS.113.0, AS2UDS.116.0, AS2UDS.412.0, AS2UDS.583.0, and AS2UDS.659.0). Inspecting their morphologies and color distributions more closely, we find that all of these sources exhibit radial \ML\ gradients, while the three sources with \rr $> 1.5$ have a cuspy stellar-mass profile ($n \geq 5$, measured on their stellar mass maps). Thus, their high ratio of \rr\ compared to the median of our sample seems to be driven by compact stellar configurations rather than a large submm size. Furthermore, the central stellar mass densities for three of those targets are the highest within our sample (exceeding $10^4 \, {\rm M_{\sun} \, pc^{-2}}$). Based on these properties, we interpret these galaxies as systems in an evolved stage, in which a central stellar mass density has already been built up due to strong central star formation. 

\begin {figure}[htb]
\centering
 \includegraphics[width=0.49\textwidth]{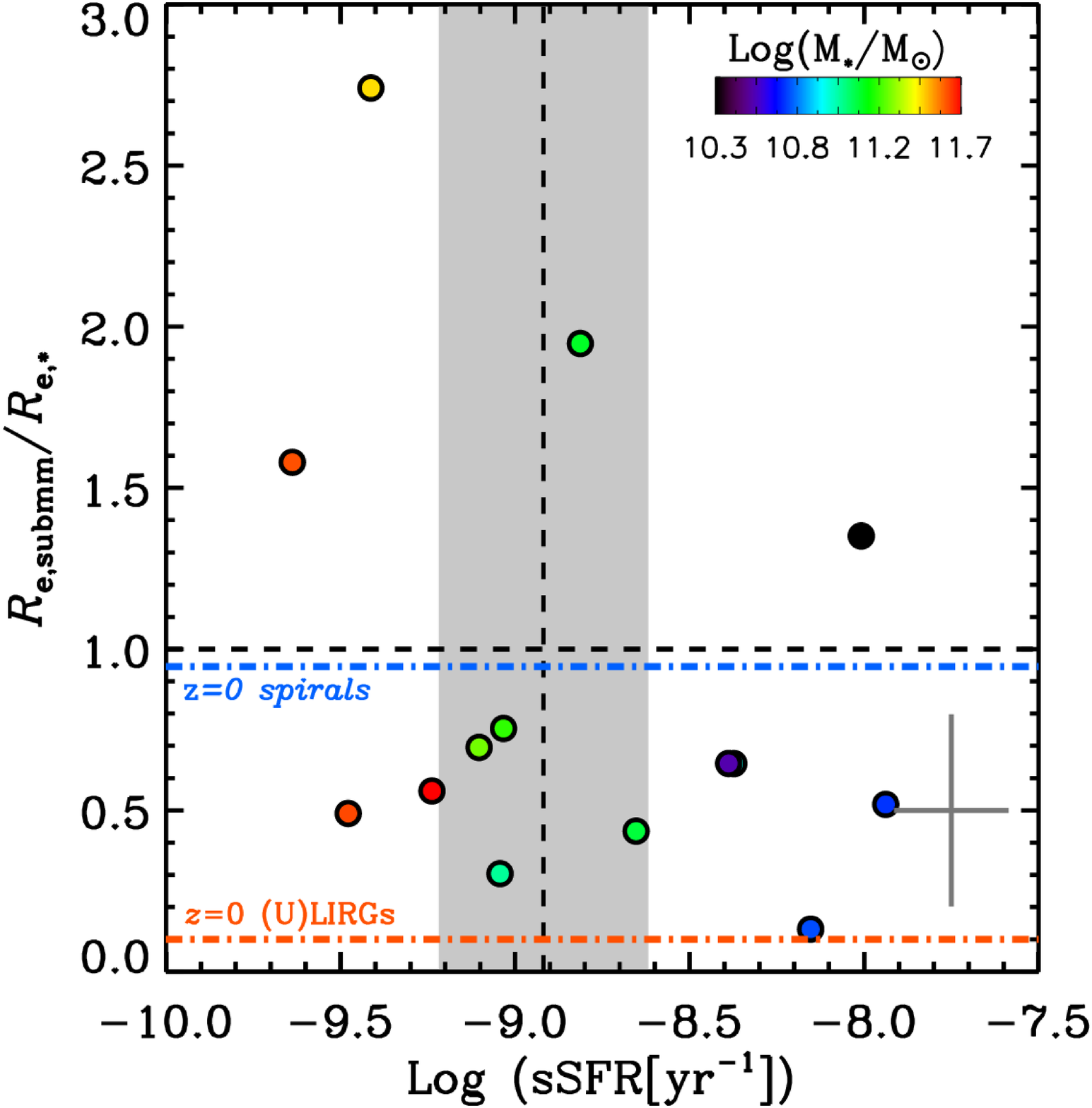}
\caption[s]{Ratio between intrinsic sizes of the submm and stellar components versus specific star-formation rate. The color-coding indicates the total stellar mass derived from the SED-modeling by {\sc Magphys}. The position and scatter of the main sequence, adopted from \cite{Whitaker2014}, is shown as the dashed vertical line and shaded area, respectively. The colored horizontal lines indicate the median size ratios for local spirals based on KINGFISH sample \citep{Hunt2015}, as well as for local (U)LIRGs as based on the GOALS survey \citep{Kim2013,Barcos2017}. Median uncertainties in the shown properties for our sample are indicated in the lower right corner. The dust to stellar size ratios show no significant correlation with the global SF properties.}
\label{Relations.fig}
 \vspace{3mm}
\end {figure}

To interpret our observations, we use the dust emission as a tracer of dust-obscured star formation in these systems, relying on the assumption that the dust properties (e.g., dust temperature) do not strongly vary spatially. This points to a picture in which the star formation is clearly more compact than the stellar distribution in high-redshift star-forming galaxies over a large range of specific star-formation rate. Recent compilations of radio size measurements of non-AGN high-redshift star-forming galaxies at 3\,GHz have shown that effective radio sizes are also smaller than the stellar component (with $R_{{\rm e,radio}} = 1.1$-$1.5$\,kpc, Jim\'enez-Andrade et al. in prep.), supporting our conclusions. For comparison to the population of local spiral galaxies, we show the median dust versus stellar size ratio of spiral galaxies measured by \cite{Hunt2015} based on the KINGFISH sample in Figure\ \ref{Relations.fig} (with \rrd\ $= 0.95$). Our SMGs exhibit a more compact dust distribution relative to the stars compare to local spiral galaxies, where the dust and stars exhibit an about equal extent.

In order to provide a reference sample for local galaxies with star-formation rate (or equivalently infrared luminosities) closer to our sample, we additionally plot the median star formation to stellar size ratio for local LIRGs and ULIRGs based on the GOALS sample. More specifically, median stellar sizes are taken from {\it HST}/ACS $i$-band measurements from \cite{Kim2013}. Due to the lack of existing dust size measurements of local (U)LIRGs with sufficient spatial resolution, we take the median effective sizes of star formation from $33 \, {\rm GHz}$ radio measurements of the GOALS sample from \cite{Barcos2017}. For computing the median \rrd\ for local (U)LIRGs, only systems with no obvious AGN contribution and/or resolved radio emission are considered. The resulting median dust versus stellar size ratio of (U)LIRGs (\rrd\ $= 0.1$) is clearly smaller than for our SMGs. We note that the available spatial resolution of star-formation tracers is significantly higher at low redshift compared to our study. However, the observed distribution of dust as a proxy for star formation in our analysis can be resolved (i.e.,\,with $R_{{\rm e,submm}} \ge R_{{e,beam}}$, where $R_{{e,beam}}$ is half of the FWHM major-axis beam size) given our high-resolution ALMA data set. One exception is the source AS2UDS.153.0 that shows a very compact dust distribution of $R_{{\rm e,submm}} = 0.67^{+0.19}_{-0.13}$\,kpc, and thus $R_{{\rm e,submm}}/R_{{\rm e,beam}} = 0.8$. This size difference demonstrates that star formation in SMGs at high redshift might be triggered differently than in the local (U)LIRG population. We discuss the implications of the morphological differences between SMGs and the local spiral and (U)LIRG populations more thoroughly in Section\ \ref{discussion.sec}.

\section{Discussion}

\subsection{Connection of SMGs to present-day ETGs}
\label{ETG.sec}

\begin{figure*}[bht]
\centering
 \includegraphics[width=0.9\textwidth]{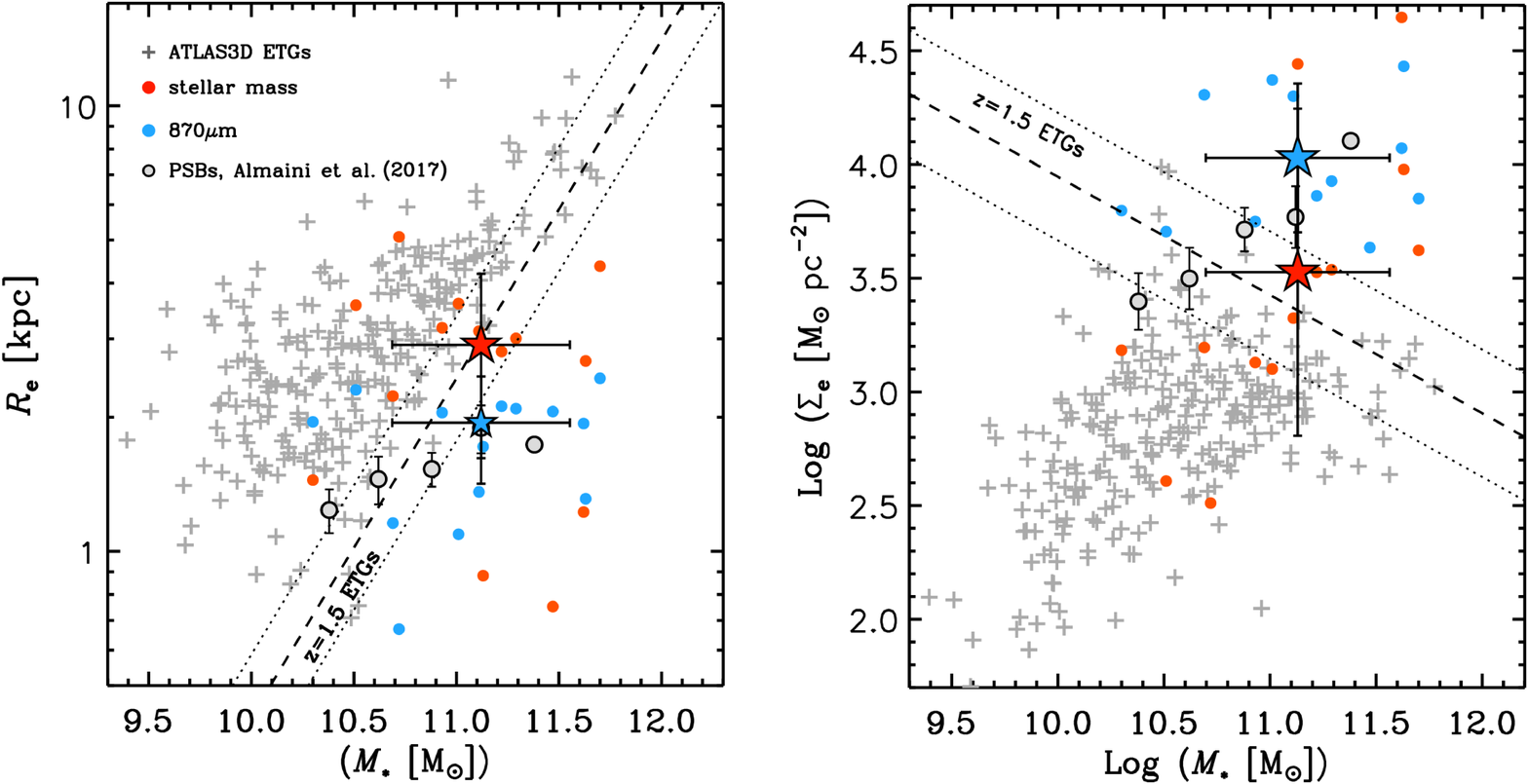}
\caption[s]{Comparison between the effective radii (left) and surface densities (right) of our SMG sample with local early-type galaxies from the \at3 survey \citep{Cappellari2011}. The measurements of the submm and stellar components for our SMGs are shown as blue and red filled circles, respectively. Median values for our entire SMGs are indicated as stars, with the error bars showing the scatter among our sample. The stellar components of the underlying population of \at3 galaxies is shown as gray crosses. The effective stellar sizes and resulting surface densities of quiescent early-types at $z=1.5$ derived by \cite{vdW2014} are shown as dashed lines, together with the associated scatter shown as dotted lines. Also shown are rest-frame optical sizes and surface densities of post-starburst galaxies (PSBs) at $1 < z < 2$ as measured by \cite{Almaini2017}.}
\label{ETGs.fig}
 \vspace{3mm}
\end {figure*}

Based on our structural measurements, we investigate in the link between SMGs and the population of passive galaxies, which are plausibly connected through the shut-down of star formation (frequently referred to as `quenching'). In Figure\ \ref{ETGs.fig}, we plot the inferred effective stellar sizes and resulting surface densities of our SMGs as a function of total stellar mass. Since our estimates of sizes and surface densities for individual SMGs are subject to substantial uncertainties, we also show the median and the scatter of our sample. As a reference sample for the passive galaxy population, we consider quiescent early-type galaxies at $z=1.5$, as those might represent the `direct' descendants of SMG once they have undergone quenching and evolved to the passive population within $\sim $1\,Gyr. Their sizes and surface densities are measured by \cite{vdW2014}, who have quantified their $R_{{\rm e}}$-$M_*$ relation based on large samples. We further consider nearby massive early-type galaxies as their potential ultimate descendants in the local Universe \citep[e.g.,\,][]{Hickox2012,Toft2012,Toft2017}. Thus, we additionally plot the effective sizes and resulting inferred surface density of local ETGs in the \at3 survey \citep{Cappellari2011}, measured through dynamical modeling by \cite{Cappellari2013}. As the ongoing central star formation in SMGs likely leads to an increase of stellar mass density (and equivalently to a decreasing stellar size), we use our ${R}_{{\rm e,submm}}$ measurements to provide an estimate on the effective star-forming sizes for each galaxy. We furthermore estimate the inferred surface densities of the star-forming component by computing the amount of cold molecular gas, $M_{{\rm gas}}$, acting as fuel for star formation. As demonstrated in the recent literature, the amount of cold gas is in very good relation with the luminosity at rest-frame wavelengths of 150-500\,\mum\ \citep[see\,][]{Scoville2014,Scoville2016,Groves2015}. We estimate $M_{{\rm gas}}$ via adopting the coefficients from Table 6 of \cite{Groves2015} for converting our ALMA 870-\mum\ flux densities into gas masses for our sources, and follow thereby the recipe of \cite{Schinnerer2016}.

The median stellar sizes and surface densities of our near-infrared bright SMGs are in good agreement with the quiescent population at $z=1.5$ at the same stellar mass. Since the star-forming component is even more compact than the stars for our SMGs, those seem to fade into systems that represent the smaller and denser part of the quiescent galaxies at $z=1.5$. 

Figure\ \ref{ETGs.fig} also plots the effective sizes and resulting surface densities of post-starburst galaxies (i.e.,\,systems selected to represent quiescent systems with very recent episodes of major star formation) at $1 < z < 2$ determined using rest-frame optical imaging by \cite{Almaini2017}. The post-starburst systems exhibit on average effective sizes of $1.5$--$2$\,kpc within the stellar-mass range sampled by our SMGs, in agreement with the aforementioned decrease in stellar size of SMGs before quenching. Thus, the post-starburst-phase might represent a link between the most immediate descendant of SMGs and the passive population at high redshift. The median stellar sizes and surface densities of our SMGs also occupy the locus of the most compact local ETGs that therefore might represent the ultimate descendants of SMG in the local Universe. However, we caution that \at3\ offers only a limited census on the structural properties of this most massive ETG population, due to the limited volume probed, which might affect this conclusion. We also note our comparison does not include any further structural evolution of SMGs before and after quenching. More specifically, if specific star-formation rate increases towards the outskirts in an `inside-out quenching' scenario -- as found by e.g.,\,\cite{Morselli2018,Tacchella2018} -- effective sizes increase before quenching, likely most pronounced for lower-mass systems. Furthermore, passive evolution of quiescent galaxies is suggested to also lead to an increase of effective size through processes such as minor merging \citep[e.g.,\,][]{Naab2009}.

\subsection{Are SMGs massive disks or mergers?}
\label{discussion.sec}

The recovered stellar morphologies of our sample of SMGs are more concentrated when converting rest-frame optical light into stellar mass (albeit with simplified assumptions on the dust geometry). The effective stellar sizes of our SMG sample are in good agreement with the effective sizes of the mass-matched star-forming galaxy population at $z=2$. Moreover, the conversion from optical light to stellar mass for our sample improves the spatial co-location of dust and stellar emission. Therefore, our findings advise caution when interpreting irregular and disturbed optical morphologies as massive SMGs undergoing major interactions. We show that dust attenuation is likely a major cause for irregular morphologies and apparent spatial decoupling between the dust (approximately tracing dust-obscured star formation) and the stellar distribution. Strong and patchy dust attenuation within SMGs (at least partially recovered by our \MLh\ correction) leads to more irregular morphologies and might partly explain a trend of an increasing fraction of interacting galaxies with $L_{\rm IR}$ \citep[e.g.,\,][]{Chen2015,Kartaltepe2012}. Thus, a significant fraction of massive SMGs at high redshift might represent systems with underlying stellar disks but appear as interacting systems as judged based on their optical morphology alone. However, this does not exclude that SMGs appearing as strongly disturbed systems are undergoing major galaxy interactions. 

Our exploration of the dependency of the dust versus stellar morphologies among systems classified as starburst galaxies and normal main sequence galaxies has shown that compact dust cores exist in both of these populations. Taking the dust emission as a proxy for dust-obscured star formation, this implies that compact and centrally concentrated star formation is a common feature in massive SMGs, irrespective of their overlap with the main sequence. Major interactions can (at least in the local universe) be attributed to large main sequence offsets \citep[e.g.,\,][]{Matteo2008}, and therefore the above findings indicate that SMGs undergoing strong interactions do not necessarily have more compact star-forming regions than the ones representing secular disks. This confirms the theoretical expectation that dense star-forming cores might be the results of both dissipative contraction during major mergers \citep{Bournaud2011} and secular inflow of gas in extended disks leading to the formation of bulges \citep{Dekel2014} at high redshift. Applying these lines of arguments to our SMG sample, we interpret our findings as suggesting that SMGs might not necessarily be indicative of major mergers based alone on rest-frame optical morphology in combination with compact submm emission \citep[e.g.,\,][]{Chen2015,Hodge2016}. To constrain the true fraction of SMGs where star formation is triggered through major galaxy interactions, additional observables such, as resolved gas kinematics, are required. We also note that the above conclusions, due to the design of our sample selection, are only based on the near-infrared bright SMGs and do not include systems with potentially strongest dust attenuation in the centers. Thus, this class of systems might represent a population with different morphological properties that allow different conclusions on the triggering mechanisms of SMGs at high redshift. 

The compactness of dust emission relative to the stars in star-forming high-redshift galaxies, not based on submm selections, has been confirmed by various studies in the literature \citep[e.g.,\,][]{Tadaki2017a,Tadaki2017b,Rujopakarn2016}. This seems to suggest that star formation in massive star-forming systems at high-redshift is more centrally concentrated compared to local star-forming galaxies, where the dust emission is distributed on similar scales as the stellar disks. This might in turn plausibly be the consequence of the elevated gas fractions observed in high-redshift star-forming galaxies, with $f_{{\rm gas}}$ reaching 40--60$\%$ at $z\simeq 2$ compared to $5\%$ at $z\simeq0$ \citep{Saintonge2017,Tacconi2010,Tacconi2013,Daddi2010,Bothwell2013,Genzel2015,Schinnerer2016,Scoville2016,Scoville2017}, leading to gas inflow due to disk instabilities \citep[e.g.,\,][]{Dekel2014}.

\section{Conclusions}

We have combined high-angular resolution ALMA submm + {\it HST}/CANDELS observations to investigate the structural properties of 20 massive ($\log{(M_*/{\rm M_{\sun}})}= 10.3$-$11.7$) SMGs at $1.7 < z < 2.6$. We have exploited multi-wavelength {\it HST}/CANDELS imaging at rest-frame optical wavelengths to derive robust color-corrected stellar mass distributions for our sources. We have carried out radial profile measurements on both ALMA imaging and stellar mass distributions, and related those to the integrated star-forming properties to shed light on the formation mechanisms of SMGs at high redshift. 

Our main results are the following. 

\begin{enumerate}
  \item By converting the $H_{160}$-band distributions of our sample into stellar mass maps using a correction for spatial \ML\ variations, we find that the stellar mass is more concentrated than inferred from single band rest-optical imaging alone. The centroid positions of stellar mass spatially coincide well with those of the dust distribution, with a median offset of $1.1$\,kpc. This is the case even for sources where the dust emission appears to be decoupled from the rest-frame optical light. 
  \item The effective sizes of our sample, as inferred from their dust emission, are on average smaller than the sizes of the stellar component ($\langle {R}_{{\rm e,submm}}/{R}_{{\rm e,mass}} \rangle = 0.6 \pm 0.2$). This size ratio does not change with integrated stellar or star formation properties, specifically with specific star-formation rate. Taking the dust emission as a proxy for dust-obscured star formation in our sources, our results imply that the SMG population at high redshift exhibits centrally concentrated star formation unlike the average population of local spiral galaxies, where star formation is as extended as the stellar distribution. 
  \item The comparison of effective stellar sizes and stellar surface densities to early-type galaxies at $z\sim1.5$ suggests that SMGs are consistent in their structural properties with fading into the passive population after the shut-down of star formation. 
 \end{enumerate}

Our study has demonstrated the importance of deriving color-corrected stellar mass maps when inferring structural properties of SMGs. We have shown that the underlying mass of SMGs is overall in better spatial agreement and closer in size compared to the dust emission than what would be inferred from rest-frame optical light. The estimated stellar sizes reveal that the dust emission is on average more compact than the existing stellar component. This indicates that the star formation in SMGs is centrally concentrated, which suggests that it is triggered through different processes than in star-forming galaxies at the present-day epoch. Finally, we show that the sizes and densities inferred from our stellar mass maps are in good agreement with early-type galaxies at $z\sim1.5$ being the descendants of SMGs at $z \simeq 2$.\\
At present, the employed color-correction when converting optical light into stellar mass might suffer from significant systematic uncertainties in the presence of high central column densities of dust in SMGs. Further observations at sufficient spatial resolution covering less extinction-sensitive infrared wavelengths, provided by e.g.,\,{\it JWST}, will be crucial to determine the underlying stellar mass of SMGs more robustly. This will ultimately allow more substantial conclusions on their origin and evolutionary connections to the local galaxy population. 

\section*{Acknowledgements} 
 
\acknowledgements
We thank the referee for constructive and insightful comments. This paper makes use of ALMA data ADS/JAO.ALMA\#2011.0.00294.S; \#2012.0.00307.S; \#2013.1.00118.S; \#2012.1.00090.S; \#2015.1.01528.S; \#2016.1.00434.S; and \#2017.1.01492.S. ALMA is a partnership of ESO (representing its member states), NSF (USA) and NINS (Japan), together with NRC (Canada), MOST and ASIAA (Taiwan), and KASI (Republic of Korea), in cooperation with the Republic of Chile. The Joint ALMA Observatory is operated by ESO, AUI/NRAO and NAOJ. This work is based on observations taken by the CANDELS Multi-Cycle Treasury Program with the NASA/ESA {\it HST}, which is operated by the Association of Universities for Research in Astronomy, Inc., under NASA contract NAS5-26555. PL, DL, and ES, acknowledge support from the European Research Council (ERC) under the European Union’s Horizon 2020 research and innovation programme (grant agreement No. 694343). IRS and AMS acknowledge support from STFC (ST/P000541/1). FB, BM, EV, KH and EFJA acknowledge support of the Collaborative Research Center 956, subproject A1, funded by the Deutsche Forschungsgemeinschaft (DFG). SL acknowledges funding from Deutsche Forschungsge-meinschaft (DFG) Grant BE 1837/13-1 r. EAC acknowledges support from the ERC Advanced Investigator Grant DUSTYGAL (321334) and STFC (ST/P000541/1). MJM acknowledges the support of the National Science Centre, Poland, through the POLONEZ grant 2015/19/P/ST9/04010; this project has received funding from the European Union's Horizon 2020 research and innovation programme under the Marie Sk{\l}odowska-Curie grant agreement No. 665778. ST acknowledges support from the ERC Consolidator Grant funding scheme (project ConTExt, grant No. 648179). The Cosmic Dawn Center is funded by the Danish National Research Foundation. EV acknowledges funding from the DFG grant BE 1837/13-1. JLW acknowledges support from an STFC Ernest Rutherford Fellowship (ST/P004784/2), and additional support from STFC (ST/P000541/1).

\appendix

\section{Details of the conversion between $J_{125}$ - $H_{160}$ and \MLh}

\subsection{Discussion of systematic uncertainties}
\label{ML_uncertainties.sec}
\begin {figure}[tb]
\centering
 \includegraphics[width=0.45\textwidth]{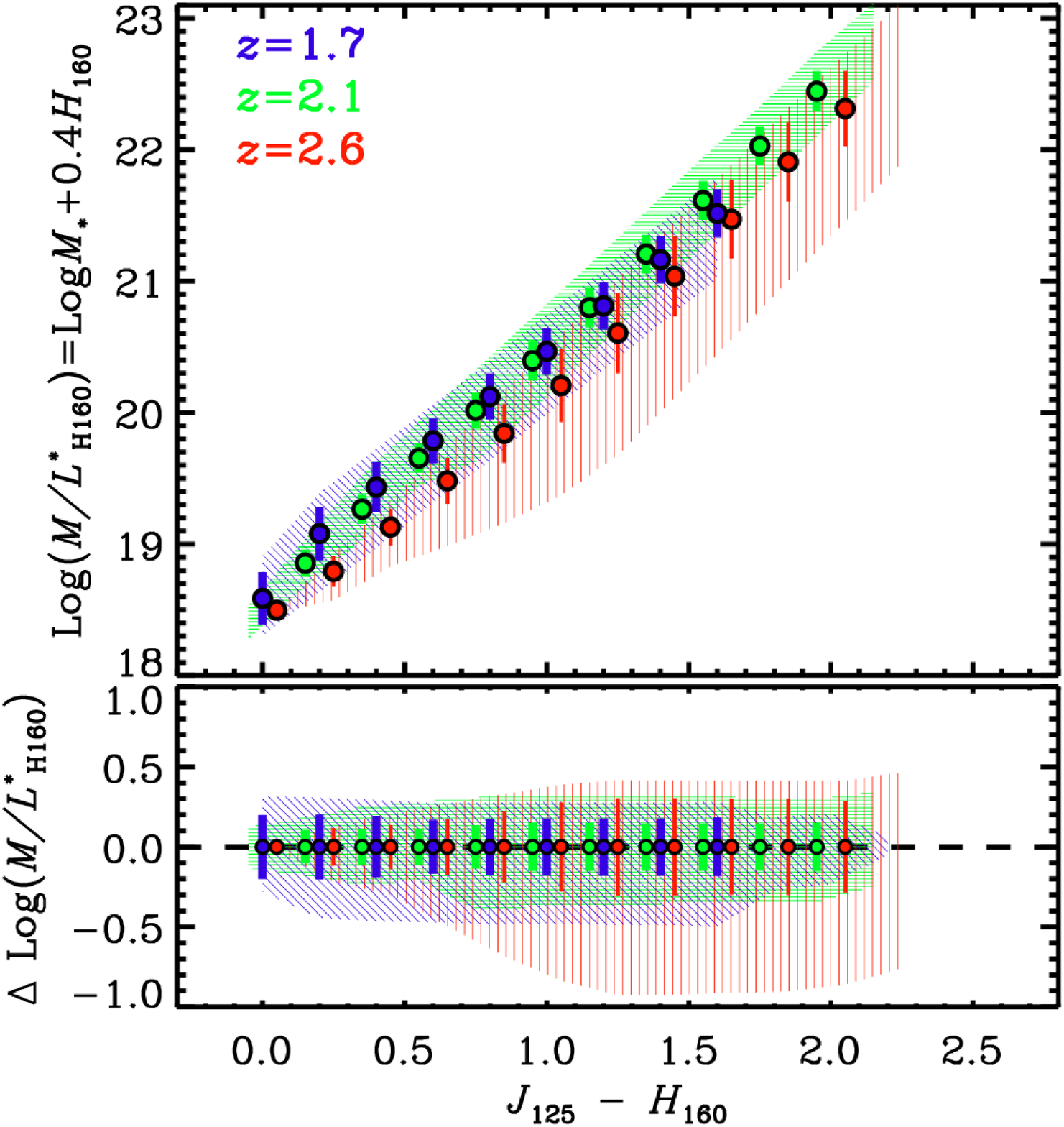}
\caption[s]{Relation between observed \MLh\ and $J_{125}$ - $H_{160}$ color for different redshifts. The top panels shows the median relations determined from our models discussed in Section\ \ref{conversion.sec}. The colored errorbars denote the scatter of \MLh\ at given color considering all models, and the polygons indicate the maximum range of \MLh\ for comparison. The bottom panel shows the relations normalized in \MLh\ to highlight the uncertainties of our relation at given observed color. }
\label{ML_polygon.fig}
 \vspace{3mm}
\end {figure}

Here, we discuss the uncertainties of our calibration of the stellar \MLh\ used to derive the stellar mass distributions, and its dependence on redshift. Figure\ \ref{ML_polygon.fig} plots our derived relation between $J_{125}$ - $H_{160}$ and \MLh\ for the redshift bins $1.7$, $2.1$ and $2.6$, which enclose the range of redshifts considered for our SMG sample. Additionally, we show the scatter in \MLh\ of our models at given color as error bars, alongside with the maximum range of \MLh\ as colored polygons. 

Within the redshift range considered, we find a well-defined relation between the $J_{125}$ - $H_{160}$ color and \MLh\ with errors that lie within 0.2-0.3 dex, depending on both color and redshift. When selecting our sample, we have considered a redshift range for which there is optimal overlap of the $J/H$-band filters with the age and extinction sensitive Balmer break, which in turn allows an accurate calibration of \MLh\ with the single observed $J_{125}$ - $H_{160}$ color. The spread of \MLh\ in the relation of 0.2-0.3 dex, however, arises due to the remaining degeneracy of stellar age and extinction in our modeling. As those are expected to spatially vary smoothly throughout our sources, which we regard these uncertainties as our error margin of \MLh\ and therefore of the stellar mass maps in our analysis. However, we note that these uncertainties might depend on the exact choice of models considered. Therefore, we inspect the maximum range of allowed \MLh\ at given color which provides a conservative estimate on the systematic uncertainties given that both star formation history and/or extinction might vary significantly between e.g., the central regions of SMGs and their outer regions. More specifically, the maximum range of \MLh\ is determined by the difference between the most `extreme' possibilities: a less-attenuated young population (note that we consider a minimum age 20 Myr in our modeling) and a more severely attenuated older population (with a maximum age that equals the age of the Universe at given redshift). We find that this systematic uncertainty shows a range of $0.3 - 0.5$ dex at $1.7 < z < 2.5$. We note that moderate color and thus inferred \MLh\ gradients of this order for our SMGs within this redshift window (i.e.,\,18 out of 19 sources) might not reflect true variations of \MLh. However, we show that the more prominent radial color trends are significant even considering such systematic uncertainties. Considering our conversion at redshifts beyond $z=2.5$, the systematic uncertainty increases to 0.9 dex, which cautions the inferred color-correction for one sources within this redshift range (AS2UDS.271.0).

\subsection{Inclusion of additional burst models}
\label{Burst.sec}

\begin {figure*}[tb]
\centering
 \includegraphics[width=0.85\textwidth]{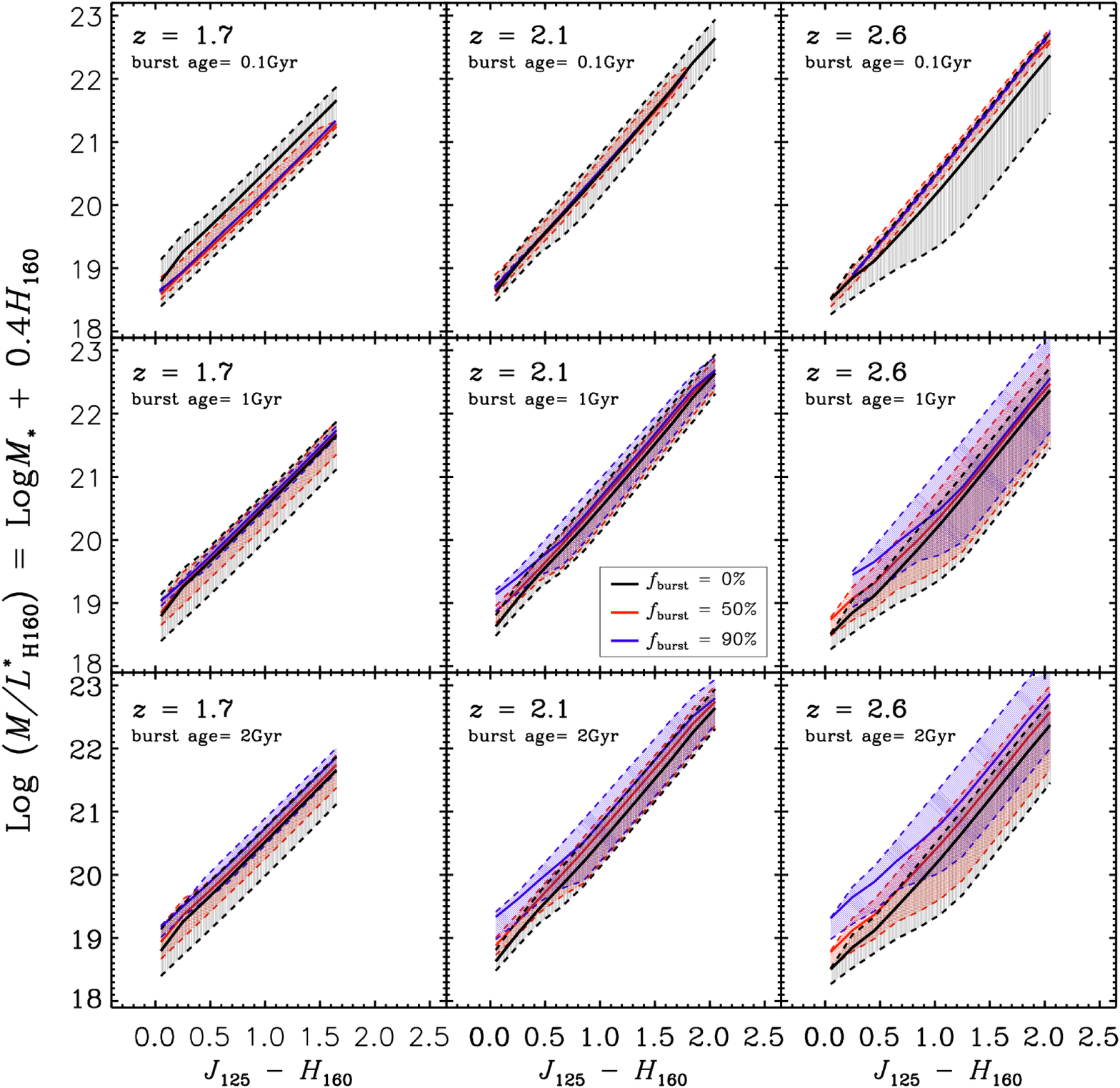}
\caption[s]{$(J_{125} - H_{160})$-\MLh\ relation including models with an additional burst of star formation. The different panels show the grid of models spanning the redshift range $1.7 < z < 2.6$, burst ages ranging from 0.1 to 2 Gyr, and burst mass fractions from $0$ to $90$\,\%. Each panel shows the median relation determined for our fiducial model without a burst (black lines), and with additional bursts with mass fractions of 50\,\% and 90\,\% (red and blue lines, respectively). The shaded polygons indicate the uncertainties based on the range of allowed \MLh\ of our models at a given color.}
\label{Burst.fig}
 \vspace{3mm}
\end {figure*}

Next, we discuss the impact of considering models with additional bursts of star formation to the derived relation between the $J_{125} - H_{160}$ color and \MLh. To do this we create a grid of additional model tracks identical to the ones described in Section\ \ref{conversion.sec}, and add a past burst event with a duration of 100\,Myr to the star-formation history. The additional bursts considered span a range of ages from 0.1 to 2\,Gyr, and a range of burst mass fractions (i.e.,\,the fraction of stellar mass produced during the burst compared to the total stellar mass of the galaxy) ranging from 0 to 90\,\%.

Figure\ \ref{Burst.fig} plots the $(J_{125} - H_{160})$-\MLh\ relation with the burst models considered in comparison to our fiducial relation without burst (i.e.,\,$f_{\rm{burst}} = 0$) for the redshift range spanned by our sample. The plotted error margins denote the systematic uncertainties discussed in the previous Section, and are based on the maximum allowed \MLh\ at a given color. 
The inclusion of past burst events leads to an overall increase of \MLh\ at a given $J_{125} - H_{160}$ color, which depends on redshift, burst strength and age. The exception is the scenario of adding a young (100\,Myr) burst component to our models at redshift 1.7, in which case the \MLh\ decreases within the systematic uncertainties. Considering moderate bursts that contribute in equal parts to the galaxy's mass as the underlying smooth population, we find that the increase of \MLh\ is $0.1$--$0.3$\,dex, and falls within the systematic uncertainties of our relation. Furthermore, the increase of \MLh\ leads to a change in only the normalization of our relation. This would imply that merely the normalization of our inferred mass maps would be most affected by past bursts, rather than the shape and sizes of the stellar mass component.

Only when considering past burst events that largely dominate the stellar mass of the galaxy at $z\geq2.1$ and are of ages $\geq 1\,$Gyr (corresponding to a formation redshift of $z\simeq$3.0--4.2 for our models at $z$=2.1 and 2.6, respectively), we find that the relation can change both in slope and normalization, exceeding the systematic uncertainties. Depending on burst mass fraction and redshift, this change can reach up to $0.8$ dex at a given $J_{125} - H_{160}$ color. Such larger offsets occur exclusively at blue colors, $J_{125} - H_{160} \lesssim 0.5$, and represent a scenario in which the past burst event makes up a large portion of stellar mass and is dominated in brightness by a young population of stars largely unattenuated. Inspecting our $J_{125} - H_{160}$ color distributions, we find that such blue colors are on average observed at larger radii ($\simeq 10\,$kpc) within our sources. However, we expect that a strong past burst, originating from a dissipative collapse of gas, resides in a compact configuration within the centers of our sources. Thus, we speculate that such situations, in which the inferred \MLh\ is strongly underestimated due to a past burst, is very unlikely to affect the \MLh\ and mass maps of our sample. Thus we conclude that the aforementioned systematic offset of \MLh\ due to more moderate bursts might lead to an additional systematic uncertainty in our relation, although it does not exceed the uncertainties already considered.

\end{document}